\documentclass[11pt,DIV=16,parskip=half]{scrartcl}
\setcapindent{0pt}

\usepackage{sfmath}  
\usepackage[font=small]{caption}
\setkomafont{captionlabel}{\bfseries\sffamily}

\usepackage{typearea}
\areaset{17cm}{26cm}
\usepackage[utf8]{inputenc}
\usepackage{titling}
\usepackage{acronym}
\usepackage[official]{eurosym}
\usepackage{listings}
\usepackage{array}
\usepackage{tabularx}
\usepackage{pgfgantt}
\usepackage[section]{placeins}
\usepackage{tabularray}
\UseTblrLibrary{booktabs}
\usepackage{enumitem}
\usepackage{amsmath}

\newcommand{\applicants}[1]{\newcommand{\theApplicants}{#1}}

\definecolor{yellow}{RGB}{252, 211, 3}
\definecolor{green}{RGB}{60,179,113}
\definecolor{blue}{RGB}{3, 165, 252}

\ganttset{canvas/.append style={fill=none, draw=black!10, line width=1pt},
    hgrid style/.style={draw=black!5, line width=.75pt},
    vgrid={*1{draw=black!10, line width=1pt}},
    title/.style={draw=none, fill=none},
    title label font=\bfseries\footnotesize,
    y unit title=0.5cm,
    title height=1,
    include title in canvas=false,
    bar label font=\mdseries\footnotesize\color{black!70},
    bar/.append style={draw=none, fill=blue, opacity=0.4},
    group/.append style={draw=none, fill=blue},
    group left shift=0,
    group right shift=0,
    group height=.5,
    group peaks tip position=0,
    y unit chart=0.5cm,
    x unit=0.7cm,
    link/.style={-latex, rounded corners=1pt, black!63},
    link label font=\footnotesize\bfseries,
    link label node/.append style={below left=-2pt and 0pt},
    milestone label font=\footnotesize,
    group label font=\footnotesize\bfseries,
}

\DeclareNewSectionCommand[
  style=section,
  counterwithin=subsubsection,
  afterskip=1.5ex plus .2ex,
  beforeskip=3.25ex plus 1ex minus .2ex,
  afterindent=false,
  level=\paragraphnumdepth,
  tocindent=10em,
  tocnumwidth=5em
]{subsubsubsection}
\RedeclareSectionCommands[
  level=\numexpr\subsubsubsectionnumdepth+1\relax,
  toclevel=\numexpr\subsubsubsectiontocdepth+1\relax,
  increaselevel
]{paragraph,subparagraph}
\RedeclareSectionCommand[
  counterwithin=subsubsubsection,
  tocindent=12em,
  tocnumwidth=6em
]{paragraph}
\RedeclareSectionCommand[
  tocindent=14em,
  tocnumwidth=7em
]{subparagraph}

\usepackage[hidelinks]{hyperref}

\RedeclareSectionCommand[
  runin=true,
  beforeskip=.1\baselineskip]{paragraph}

\usepackage[defernumbers=true,backend=biber,sorting=none,isbn=false,url=false,maxbibnames=999,maxcitenames=2,doi=false,giveninits=true,sortcites=true,citestyle=numeric-comp,eprint=false,hyperref=true]{biblatex}

\AtBeginBibliography{\footnotesize}
\AtEveryBibitem{%
  \clearfield{note}%
}
\AtEveryBibitem{\clearfield{month}}
\AtEveryBibitem{\clearfield{day}}

\preto\fullcite{\AtNextCite{\defcounter{maxnames}{99}}}

\makeatletter
\def\nhblx@bibfile@name{\jobname -nhblx.bib}
\newwrite\nhblx@bibfile
\immediate\openout\nhblx@bibfile=\nhblx@bibfile@name

\immediate\write\nhblx@bibfile{%
  @comment{Auto-generated file}\blx@nl}

\newcounter{nhblx@name}
\newcommand*{\nhblx@writenametobib}[1]{%
  \stepcounter{nhblx@name}%
  \edef\nhblx@tmp@nocite{%
    \noexpand\AfterPreamble{%
      \noexpand\setbox0\noexpand\vbox{%
        \noexpand\nhblx@getmethehash{nhblx@name@\the\value{nhblx@name}}}}%
  }%
  \nhblx@tmp@nocite
  \immediate\write\nhblx@bibfile{%
    @misc{nhblx@name@\the\value{nhblx@name}, author = {\unexpanded{#1}}, %
          options = {dataonly=true},}%
  }%
}

\AtEndDocument{%
  \closeout\nhblx@bibfile}

\newcommand*{\nhblx@boldhashes}{}
\DeclareNameFormat{nhblx@hashextract}{%
  \xifinlist{\thefield{hash}}{\nhblx@boldhashes}
    {}
    {\listxadd{\nhblx@boldhashes}{\thefield{hash}}}}

\DeclareCiteCommand{\nhblx@getmethehash}
  {}
  {\printnames[nhblx@hashextract][1-999]{author}}
  {}
  {}

\newcommand*{\addboldbibnames}{\forcsvlist\nhblx@writenametobib}
\newcommand*{\resetboldbibnames}{\def\nhblx@boldhashes{}}

\newcommand*{\mkboldifhashinlist}[1]{%
  \xifinlist{\thefield{hash}}{\nhblx@boldhashes}
    {\mkbibbold{#1}}
    {#1}}
\makeatother

\DeclareNameWrapperFormat{boldifhashinlist}{%
  #1}

\DeclareNameWrapperAlias{sortname}{default}
\DeclareNameWrapperAlias{default}{boldifhashinlist}

\DeclareNameFormat{FirstLast}{%
  \ifthenelse{\value{listtotal}=1}
    {\namepartgiveni\adddot\addspace\ifblank{\namepartprefix}{}{\namepartprefix\addnbspace}\namepartfamily}
    {\ifthenelse{\value{listtotal}=2}
      {\ifthenelse{\value{listcount}<\value{listtotal}}
        {\namepartgiveni\adddot\addspace\ifblank{\namepartprefix}{}{\namepartprefix\addnbspace}\namepartfamily\addspace}
        {\bibstring{and}\addspace\namepartgiveni\adddot\addspace\ifblank{\namepartprefix}{}{\namepartprefix\addnbspace}\namepartfamily}}
      {\ifthenelse{\value{listcount}=1}
        {\namepartgiveni\adddot\addspace\ifblank{\namepartprefix}{}{\namepartprefix\addnbspace}\namepartfamily\addcomma\addspace\dots\isdot\addcomma}
        {\ifthenelse{\value{listcount}<\value{listtotal}}
          {}
          {\addspace\namepartgiveni\adddot\addspace\ifblank{\namepartprefix}{}{\namepartprefix\addnbspace}\namepartfamily}}%
      }%
    }%
}

\DeclareRefcontext{own}{labelprefix=PR}
\DeclareRefcontext{ownpreprint}{labelprefix=PRP}

\DefTblrTemplate{caption-tag}{nocaption}{}
\SetTblrTemplate{caption-tag}{nocaption}
\DefTblrTemplate{caption-sep}{nocaption}{}
\SetTblrTemplate{caption-sep}{nocaption}

\usepackage{amssymb}
\usepackage{bbm}
\usepackage{tikz}
\usetikzlibrary{bayesnet}
\usetikzlibrary{positioning}
\applicants{Dr. Who, University of Foo}
\title{My fancy project}
\addboldbibnames{{Weiss, Martina},{Köster, Johannes}}

\begin{document}

\textbf{\huge Assessing the Viability of Wave Field Synthesis in VR-Based Cognitive Research}

\vspace{1em}
\hrule
\vspace{1em}

\today

\textbf{Benjamin Kahl - Max Planck Institute for Human Development}
\vspace{1em}

\section{Introduction}

In recent years, the use of virtual reality (VR) in cognitive and behavioral science has grown significantly \cite{KaiyanZhu}, with an increasing number of publications highlighting its advantages in terms of ecological validity, reproducibility, and experimental flexibility.

\begin{figure}[h]
    \centering
    \includegraphics[scale=0.36]{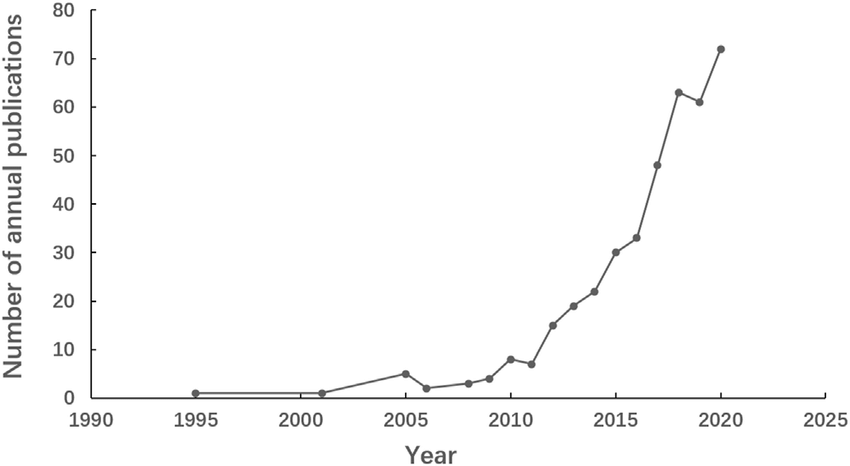}
    \caption{The number of annual publications on VR application in the MCI population published from 1995 to 2020 in Web of Science. Figure by Zhu et al.\cite{KaiyanZhu} under CC BY-NC-ND 4.0.}
    \label{fig:vr-pubs}
\end{figure}

VR allows researchers to create controlled yet dynamic environments, making it an attractive tool for studying human perception, decision-making, and behavior \cite{Faria2023, Parsons2015}. However, a recurring challenge is determining how well findings from VR-based studies generalize to real-world settings. The term of \textit{ecological validity} has thus become recurring in the psychology-vr space, and researchers hope to see an increase in immersion by integrating other senses into their virtual experiments. While most studies focus heavily on visual stimuli, auditory cues are often underutilized despite their critical role in perception and spatial awareness \cite{Bosman02012024}.

The ability of participants to move freely within VR environments presents a unique opportunity to leverage spatially accurate soundscapes, such as those generated through artificial wavefield synthesis, to create a more immersive and realistic experimental setting.

\subsection{Wave-Field Synthesis}

\begin{figure}[h]
    \centering
    \includegraphics[scale=0.3]{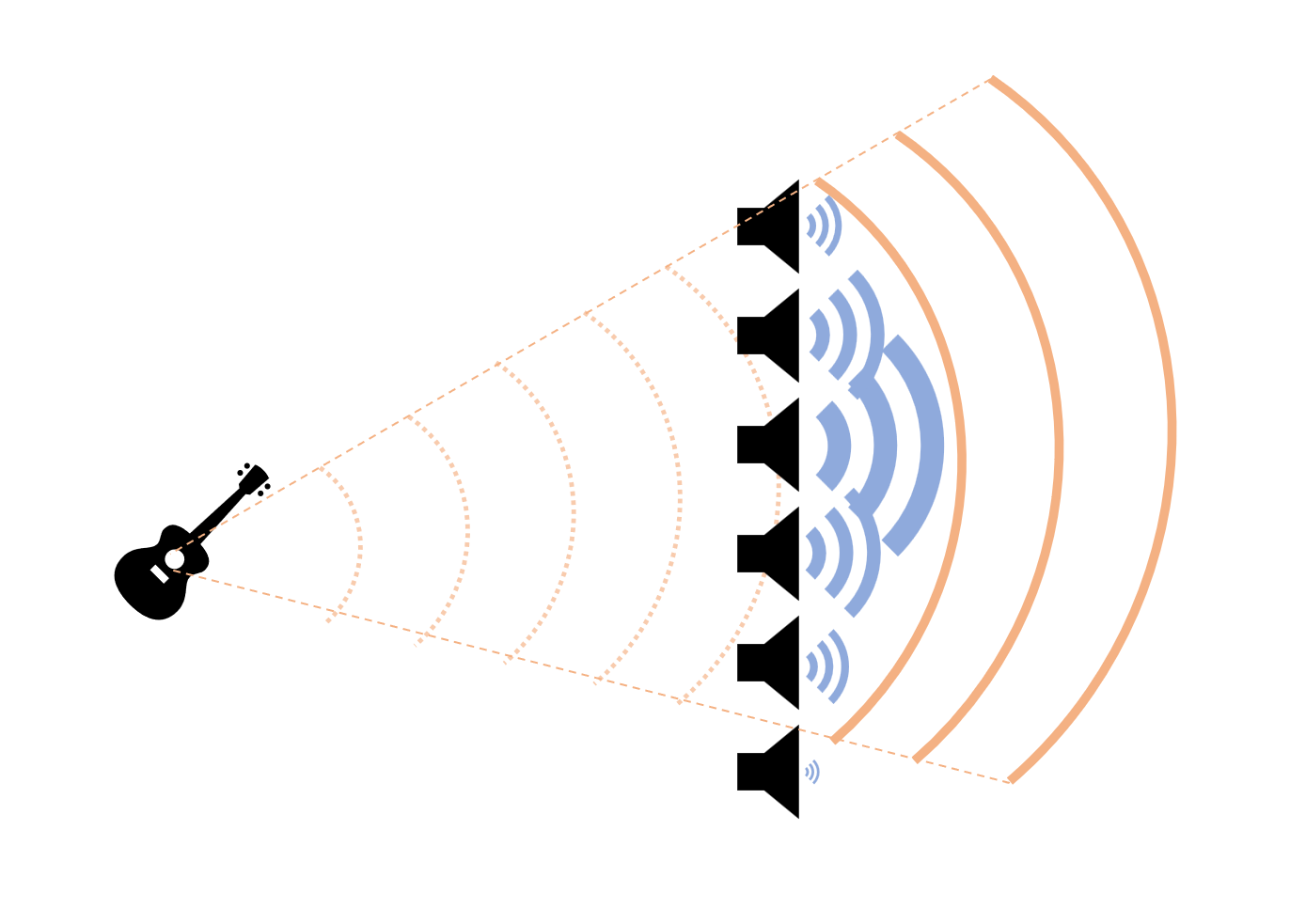}
    \caption{The concept behind wave field synthesis: Each speaker produces an elementary sound-wave which, in conjunction, produce a combined soundwave that approximates the sound of an arbitrary emitter.}
    \label{fig:wfs-concept}
\end{figure}

Wave field synthesis (WFS) is an audio rendering technique for simulating soundwaves conceived by Berkhout in 1993 \cite{Berkhout1993} and refined by Brandenburg et al. in 2009 \cite{WFS_og}. As illustrated in fig. \ref{fig:wfs-concept}, loudspeaker arrays synthesize artificial wavefronts that give the illusion of originating from a given start-point by super-imposing a large set of individual, elementary sound-waves. \cite{WFS_og}

These systems allow for the precise placement of sound-sources and allow listeners to experience the artificial soundscapes through their natural sense of hearing, thus providing an unprecedented amount of auditory immersion.

Due to their high cost, complexity and sensitivity to room acoustics, WFS systems are hitherto seldom used and have not yet reached market maturity. This novelty and limited accessibility opens up a new frontier of research thus far untrodden in the field of psychology.

Unlike commonly available, software-based spatial audio solutions such as \textit{Steam Audio}\footnote{\url{https://valvesoftware.github.io/steam-audio/}} or \textit{Dolby Atmos}\footnote{\url{https://www.dolby.com/technologies/dolby-atmos/}}, WFS-based systems do not rely on HRTFs.
As such, WFS systems are agnostic to the listener and circumvent the inter-subject variability that inhibit the reliability of HRTF-based systems \cite{So2010} in cognitive and behavioral studies.

\subsection{Goals}

This project aims to evaluate the viability of using Wavefield Synthesis (WFS) for VR-based studies, identifying its challenges and limitations, as well as their potential solutions.

To achieve this, we will implement and conduct a sample study that examines the accuracy of perceiving sound origins. Participants will be tasked with locating sound sources in both a WFS-rendered environment and a conventional stereo-headphone setup, allowing for a direct comparison of localization performance.

Additionally, VR will be leveraged to investigate how different virtual environments, sound types, and other experimental parameters influence both accuracy and search behavior.

All in all, we hope to produce a comprehensive evaluation of both the viability of WFS in psychology, as well as an early assessment for the perceived accuracy of sound localization such systems provide when compared to more traditional approaches.

\section{Technology and Setup}

\subsection{WFS Setup}

\begin{figure}[h]
    \centering
    \includegraphics[scale=0.12]{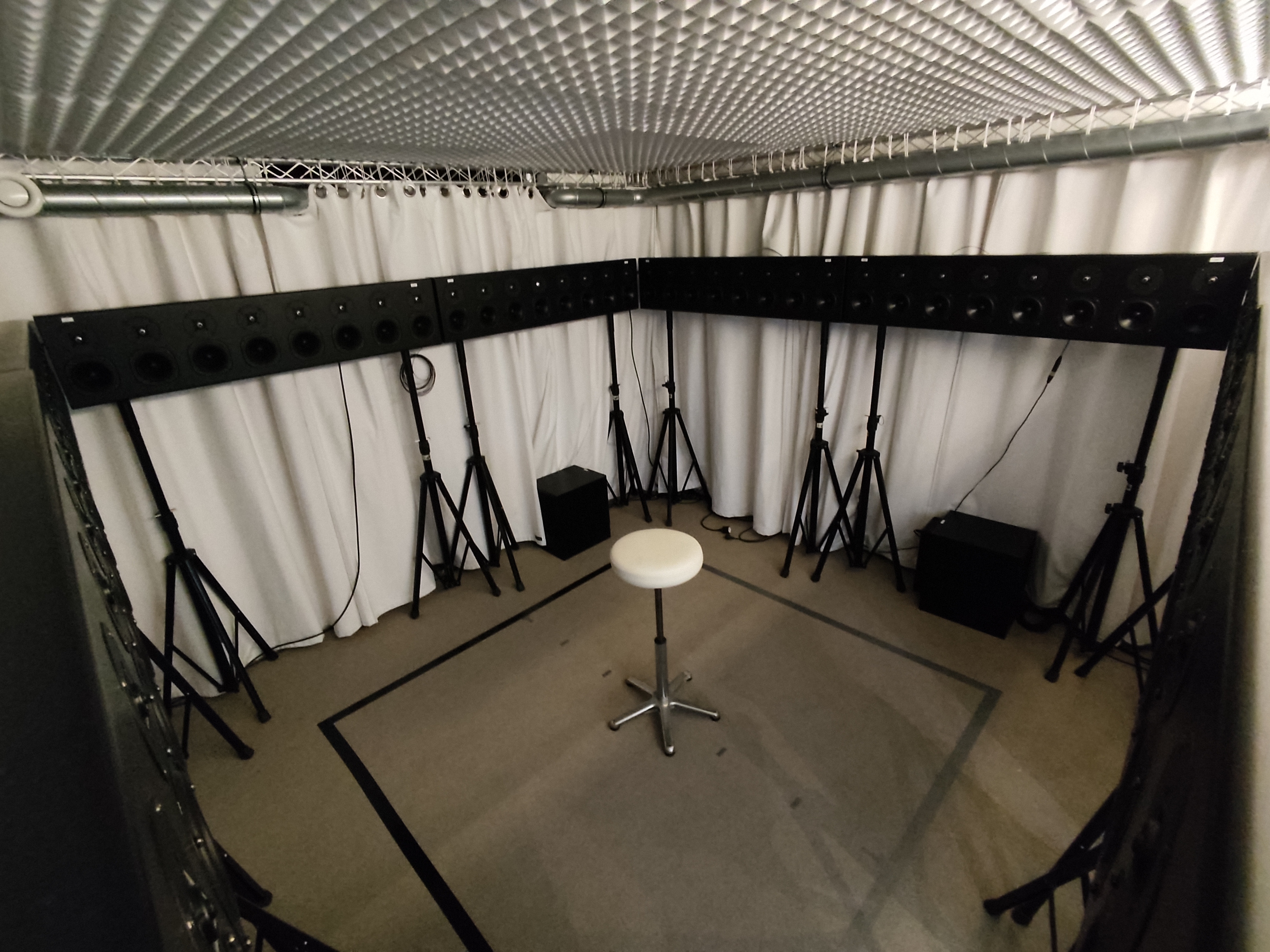}
    \caption{The employed WFS setup at the MPIB. The area marked in black tape corresponds to the WFS-area the participant can move around in.}
    \label{fig:wfs-setup-mpib}
\end{figure}

The Wavefield Synthesis system used throughout this project (shown in fig. \ref{fig:wfs-setup-mpib}) was developed, installed, and maintained by Fraunhofer IDMT\footnote{\url{https://www.idmt.fraunhofer.de/}} for use at the Max Planck Institute for Human Development\footnote{\url{https://www.mpib-berlin.mpg.de/en}} (MPIB). The setup consisted of a sound-isolated room housing a dedicated sound-rendering PC and a square-shaped WFS speaker array. The system featured four linear arrays of 16 speakers each, forming a square configuration capable of synthesizing artificial sound fields within a 2×2 meter interaction zone. All speakers were positioned at head height, meaning that virtual sound sources could only be placed within the horizontal plane parallel to the floor. Additionally, three subwoofers on the ground provided complementary low-frequency support.

\begin{figure}[h]
    \centering
    \includegraphics[scale=0.3]{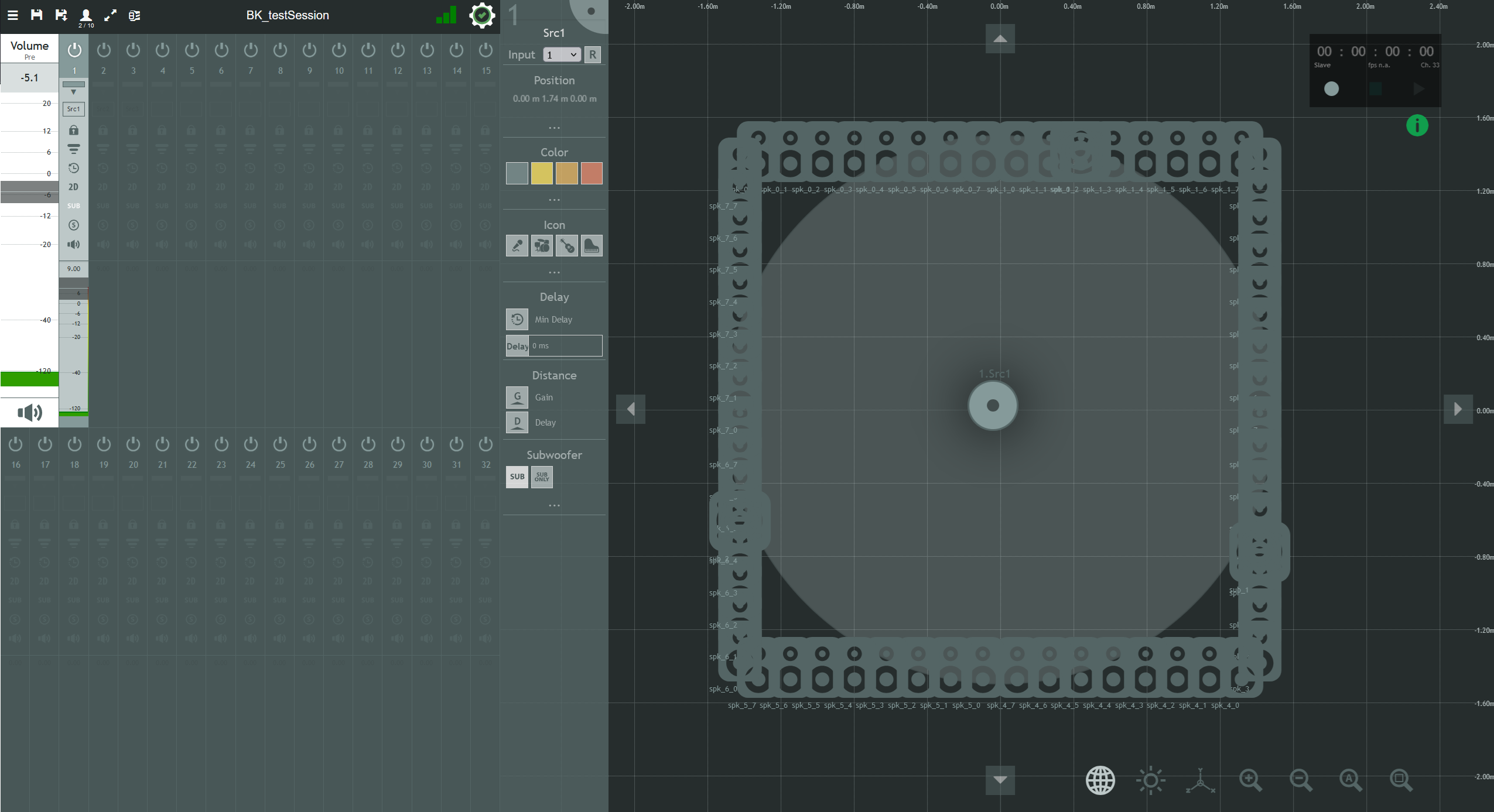}
    \caption{Fraunhofer's SpatialAudio web-interface allowing the WFS system to be configured. We used a single point-like sound source and set it's position by sending OSC command messages to the render PC from out Unity applications (see section \ref{docs:osc-api}).}
    \label{fig:wfs-interface}
\end{figure}

Audio playback was facilitated via a USB connection to the rendering PC, utilizing the appropriate driver setup through MADIFace\footnote{\url{https://rme-audio.de/madiface-usb.html}}. Most fixed system parameters were configured in advance using Fraunhofer's SpatialAudio web interface (see fig. \ref{fig:wfs-interface}), while real-time control over sound source positioning was achieved by sending UDP packets containing OSC messages to the rendering PC (see section \ref{docs:osc-api} for further details). This setup ensured precise spatial audio rendering and allowed for dynamic adaptation of sound source locations during the experiment.

\subsection{VR Setup}

We opted to use a Meta Quest Pro\footnote{\url{https://www.meta.com/de/en/quest/quest-pro/}} as our VR headset, due to its wireless capabilities and built-in hand-tracking, minimizing physical constraints and allowing participants to move freely within the virtual environment and soundscape. Rather than developing a standalone application to run directly on the headset, we utilized Meta's Air Link feature to stream the VR experience from a PC to the headset over a high-speed wireless connection.

To ensure a stable and low-latency VR stream, we deployed a 5GHz Wi-Fi 6 router inside the WFS room, positioned in direct view and close proximity to the headset. The Unity application powering the VR experience ran on a Lenovo Legion 5 Pro equipped with an Nvidia GeForce RTX 3070 Ti (Laptop). This intermediary PC managed the VR rendering and streaming while also serving as the bridge to the WFS system. It was connected to the WFS rendering PC via both USB and a dedicated Ethernet connection, using USB for audio playback and Ethernet for real-time sound source positioning.

\begin{figure}[h]
    \centering
    \includegraphics[scale=0.15]{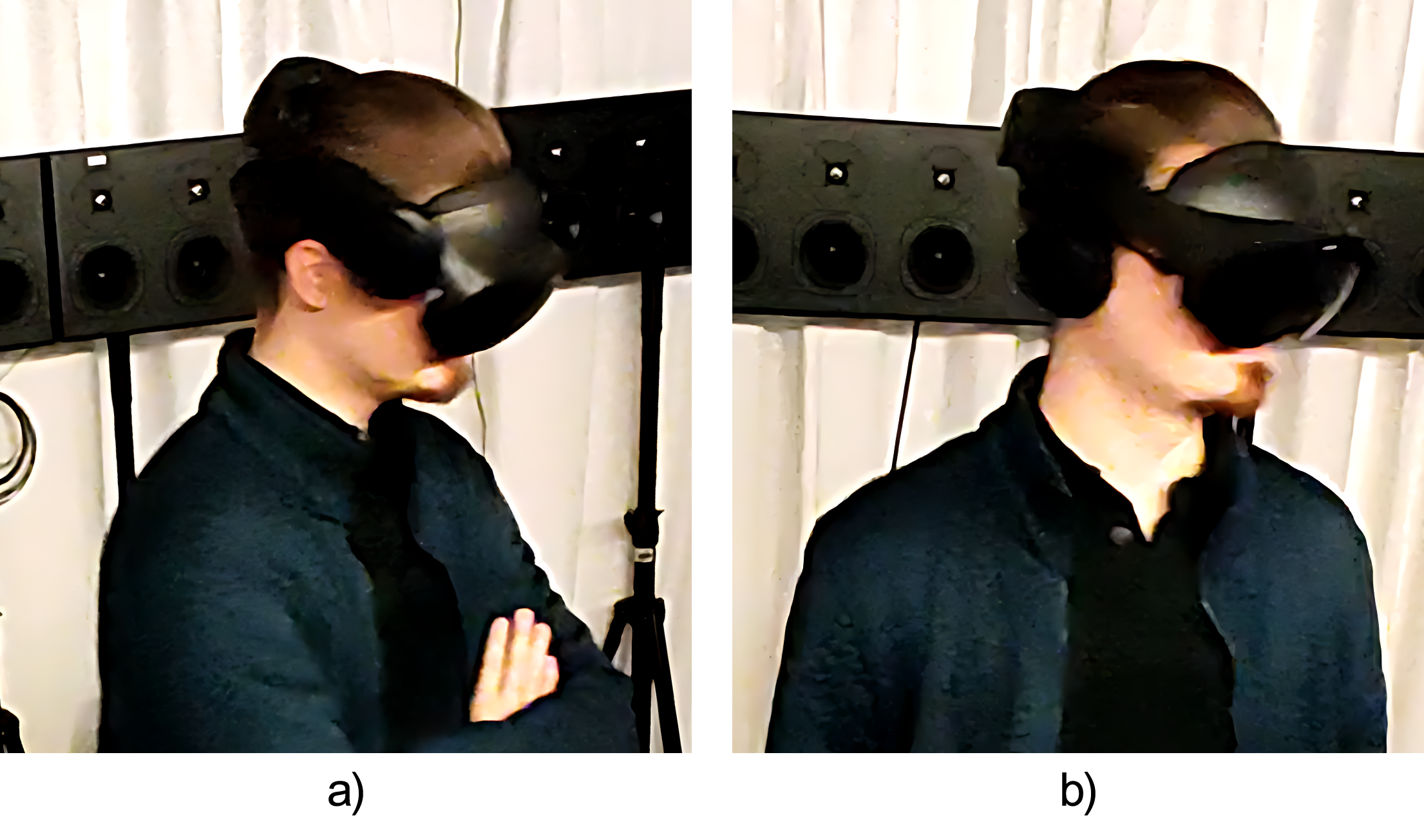}
    \caption{Participant wearing a Meta Quest Pro during WFS trials (a), and stereo trials (b). During WFS trials, the stereo headphones would be folded up and sound would come from the surrounding WFS speakers.}
    \label{fig:headphones}
\end{figure}

For the stereo headphone condition, we used a pair of Quest Pro-compatible attachable stereo headphones from Globular Cluster\footnote{\url{https://www.globular-cluster.com/meta-quest-pro-headphones.html}}, ensuring consistent and high-quality audio output for comparison with the WFS-rendered soundscape. A participant wearing the headphone attachments can be seen in fig. \ref{fig:headphones}.

\subsection{Software Setup}

The project was developed using Unity 2021.3.17, integrating the extOSC add-on to facilitate communication with the WFS renderer via its OSC interface. Most of the VR components were built using \textit{ARC-VR}\footnote{\url{https://github.com/MPIB/arc-vr}}, an in-house framework developed at MPIB for VR-based research.

For experiments utilizing individual sound sources, extOSC provided a seamless and straightforward integration with the WFS system, allowing real-time control over sound positioning. However, handling multiple simultaneous audio sources may introduce additional complexity, which has not yet been tested or implemented in this study.

To switch between audio output devices (WFS and stereo headphones) dynamically within Unity, we utilized \textit{SVCL}\footnote{\url{https://www.nirsoft.net/utils/sound_volume_command_line.html}}, a command-line tool by Nirsoft. Additionally, voice instructions for tutorial sessions and transition announcements between experiment blocks were generated using \textit{ElevenLabs}\footnote{\url{https://elevenlabs.io/}}. A custom Unity plugin was developed to allow direct generation of these voice files within the Unity environment. A more detailed account of our Unity application is provided in section \ref{tech-documentation}.

Models and sound assets were sourced from the Unity Asset Store, CGTrader.com, and Freesound.org, ensuring a diverse selection of experimental stimuli. Data analysis was conducted using a custom Python framework tailored to the needs of the study.

\begin{figure}[h]
    \centering
    \includegraphics[scale=0.45]{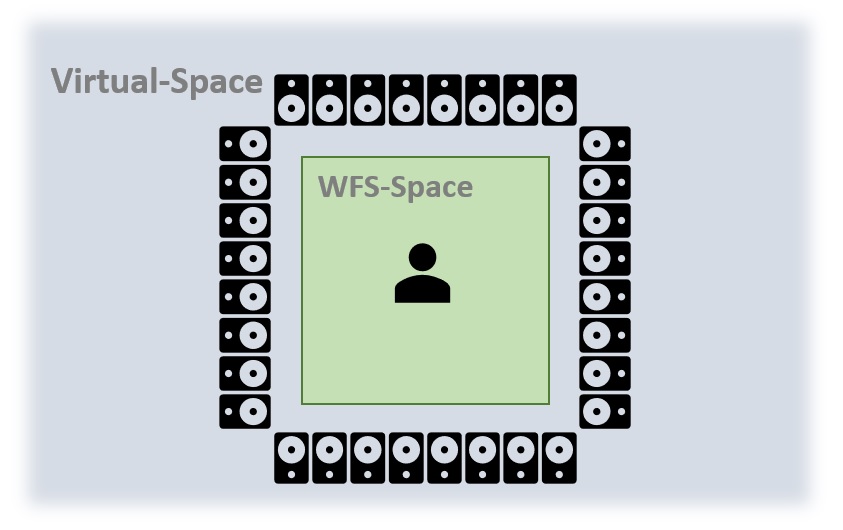}
    \caption{Difference between WFS-space and virtual-space: Whilst the participant could only move around and hear the artificial wave-fronts inside a $2\times2$m area, the size of the virtual environment (where virtual objects sound sources could be placed) extended far beyond.}
    \label{fig:wfs-setup}
\end{figure}
\section{Development}

Development of the study followed an iterative model in which we tested different prototypes and adjusted our design based on observations and feedback. This phase of exploration was primarily centered around different user-input methods in selecting/guessing a sound location.

Below we give a brief overview of each major prototype tested, as well as the related observations. In section \ref{final-version-ch} we describe the last version used in data collection and how it addresses the shortcoming observed in earlier prototypes.

\subsection{Initial Prototype}

As a first step, a simple prototype was developed to test the feasibility of using WFS in a VR-based localization task. The prototype was set in a virtual bar-like environment measuring 10×10 meters, populated with telephones placed at arbitrary locations within the space.

During the task, one of the phones would emit a ringing sound every 3–10 seconds, prompting the participant to identify its source using a virtual laser-pointer selection system. If the correct phone was selected, it would disappear. The trial continued until all phones were removed from the room.

\begin{figure}[h]
    \centering
    \includegraphics[scale=0.35]{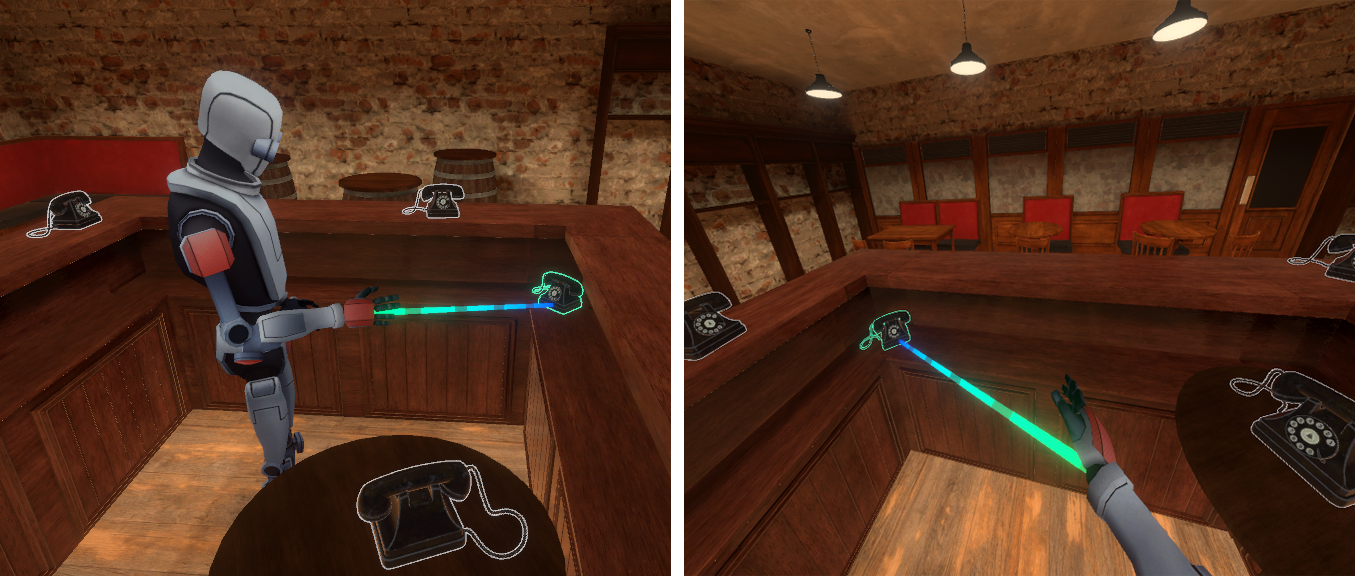}
    \caption{Participant in virtual space (visualized by a robot) pointing out a phone. Third person view (left) and first-person view (right).}
    \label{fig:first_proto_1}
\end{figure}

While the prototype demonstrated promise and was generally perceived as engaging, it also revealed several limitations. The most significant issue stemmed from the discrete selection process, where the participant was limited to choosing from a fixed number of phones. This constraint meant that even with poor localization accuracy, a participant could still identify the correct phone purely by chance, especially if only a few phones remained. Conversely, even with highly accurate localization, participants might still make incorrect selections in trials with a large number of phones simply due to minor perceptual errors.

As a result, this arbitrary limitation introduced noise into any qualitative measurements of localization accuracy, making it clear that a more refined experimental design would be necessary for meaningful evaluation.

\begin{figure}[h]
    \centering
    \includegraphics[scale=0.38]{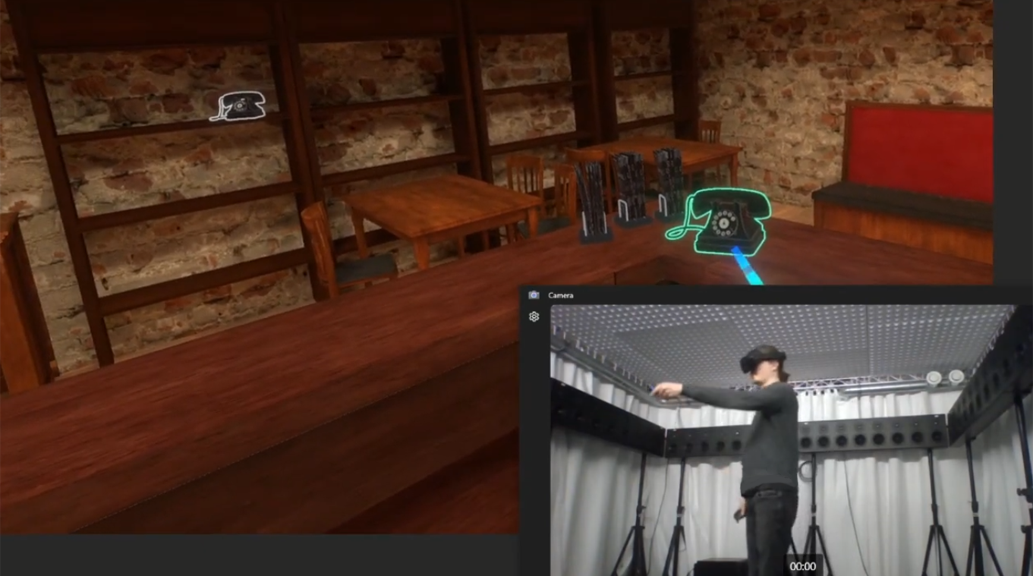}
    \caption{Participant inside the WFS system selecting a phone. The VR view is streamed over WIFI to a \textit{Meta Quest Pro} headset and presented to the participant. For a full video-demonstration, refer to the following link: \url{https://youtu.be/CWmqsejxNqI}.}
    \label{fig:first-proto}
\end{figure}

As a follow-up to the initial prototype, we devised a set of four re-designs that, hopefully, allow for a reliable measurement of an euclidean-distance \textit{delta-value} between the location the participant perceives/indicates, and the location that the wavefront was actually created at.

\subsection{Phone-Placement Version}
\label{phone-placer-proto}

In this variant, the participant did not \textit{choose} one of several options, but instead pointed out the location they believed the sound to have originated from directly.

After a ring was sounded, the participant's VR controller was given a 'virtual phone' which they could move and place into a desired location using the controller's trigger button.

\begin{figure}[h]
    \centering
    \includegraphics[scale=0.3]{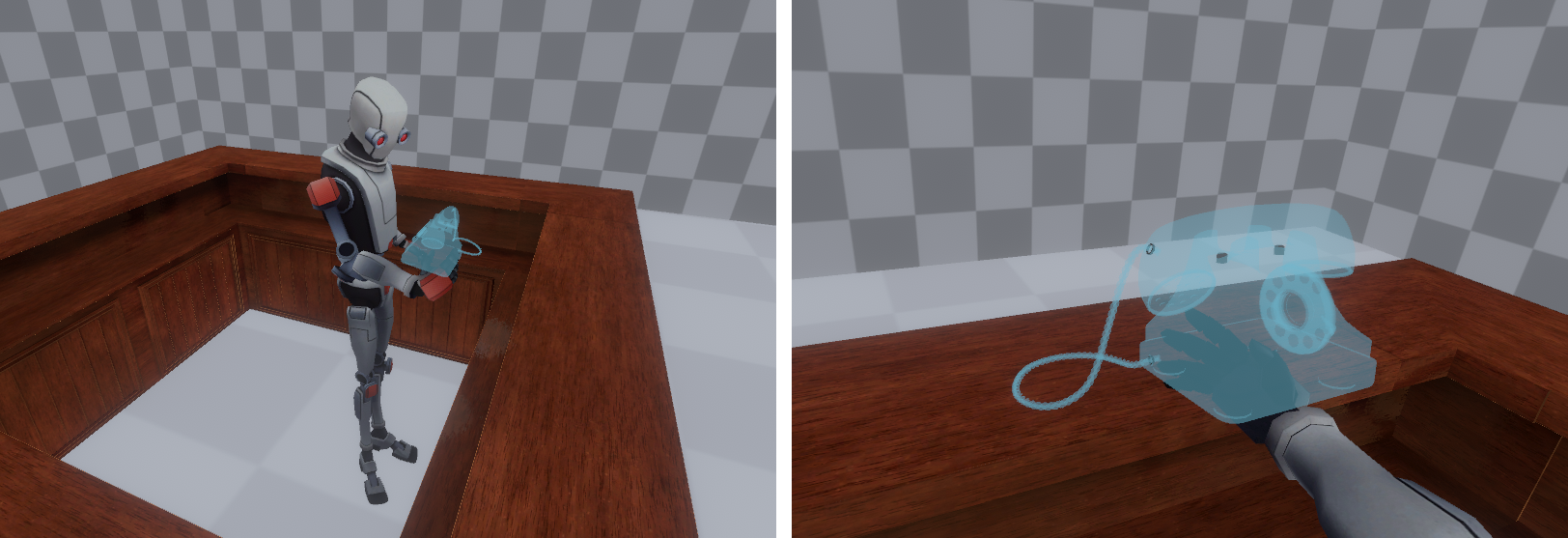}
    \caption{The participant in virtual space, represented by robot, placing a virtual phone at the location of the perceived sound. Third-person view (left) and from first-person view (right)}
    \label{fig:enter-label}
\end{figure}

\newpage

Observations:

\begin{itemize}
    \item Due to being tied to the exact location of a participant's hand/controller, phones could not be placed outside of the limited WFS-space, thus all sounds had to originate from within as well.
    \item The participant could place phones above or below the height of the speakers, even though the used WFS system could only localize sounds at one height.
    \item This version provided intuitive and simple controls for inexperienced VR users, but could benefit even further from substituting the controllers by hand-tracking.
\end{itemize}

\subsection{Dot-Circle Version}
\label{dot-circle-proto}

In this variant the participant was provided an array of dots in a circle at head-height around the center of the WFS-space, of which they could select the one they believed to be closest to the sound's origin.

\begin{figure}[h]
    \centering
    \includegraphics[scale=0.28]{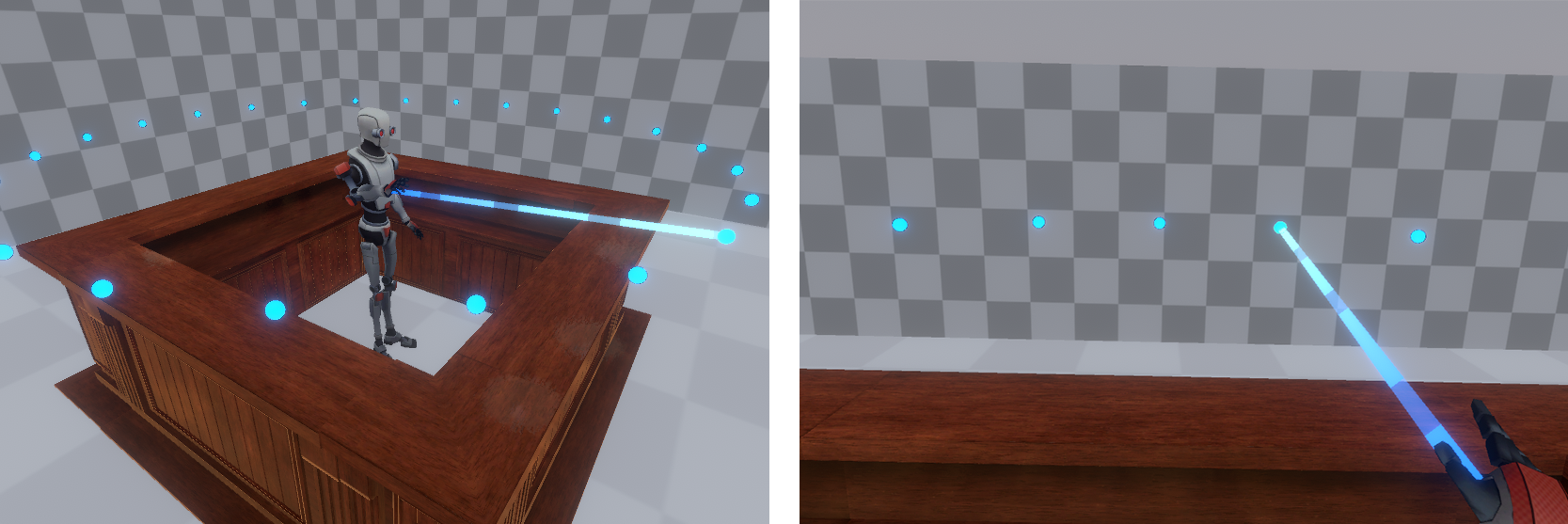}
    \caption{The participant in virtual space, represented by robot, pointing out the sound's location from a circle of options. Third-person view (left) and from first-person view (right)}
    \label{fig:enter-label}
\end{figure}

Observations:

\begin{itemize}
    \item This more abstract variant made no reference at all to 'telephones' or other explicit sound-sources. Whilst less immersive, it could thus work better with entirely different sounds.
    \item Adjusting the delta-angle of the placed dots to choose from allowed for the measurement of a discrete distance/angle at which the accuracy of localization begins to vein.
    \item This version allowed sounds to originate from outside the WFS area, but since all the dots are located on a circle around the room's center, the participant does not indicate the 'depth' at which the sound originated from, only the direction. A depth-selection would either require one of the other variants, or limiting the phone-distance to a single value per trial.
    \item Since the primary input parameter was measured as an \textit{angle} instead of a distance, this variant would provide a counter-balance to the expected inverse-square decrease in accuracy.
    \item This variant dis-incentivized the participant from moving around the virtual space, as the parallax-effect of the sound origin (see fig. \ref{fig:sound-parallax}) is no longer as useful.
\end{itemize}

\subsection{Laser-Pointer}

In this variant, the participant could point out a location on a virtual 'canvas' at head-height by using a virtual laser-pointer.

\begin{figure}[h]
    \centering
    \includegraphics[scale=0.3]{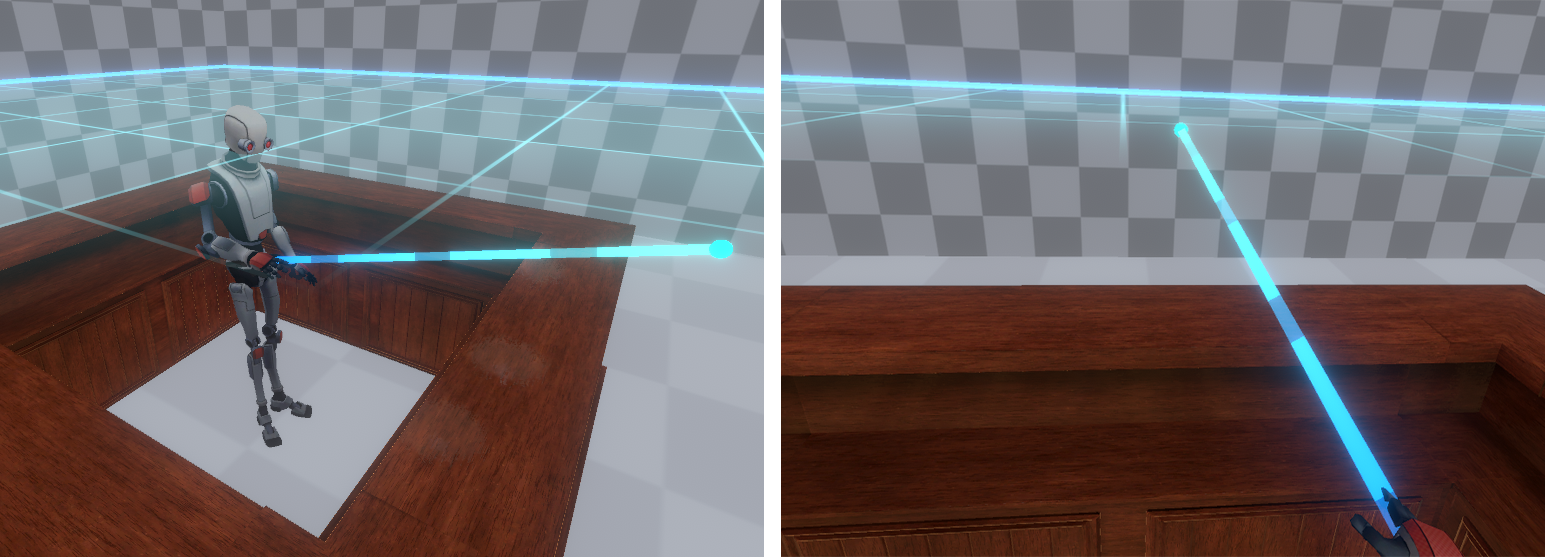}
    \caption{The participant in virtual space, represented by robot, pointing out the sound's location on the selection-canvas from a third-person view (left) and from first-person view (right)}
    \label{fig:enter-label}
\end{figure}

Observations:

\begin{itemize}
    \item This variant allowed participants to both place phones inside and outside the WFS area. As such, it provided the highest resolution/options for positions out of all tested versions.
    \item The accuracy of a selected location scaled to inverse square of the distance between the participant and the location and thus leads to a natural increase in the expected delta-values.
    In other words, noise from tracking inaccuracies or the participant's hand movements were amplified the larger the selected location is, due to the increased length of the laser-ray.
    \item Having a laser-canvas on the XZ-plane at head-height was unintuitive and could provide an accuracy advantage to participants that duck or crouch.
    \item The selection plane being close to eye-height could make it difficult to see and select a location. A conceivable solution would be to move the selection plane up or down from the actual height of the speakers.
\end{itemize}

\subsection{Volume-Selection}

Testing of the above listed prototypes showed that the Phone-placement version, despite not being able to select outside the WFS area, was the most intuitive and interactive one. We built on top of this concept with a version where participants could select a certain volume from a grid.

\begin{figure}[h]
    \centering
    \includegraphics[scale=0.3]{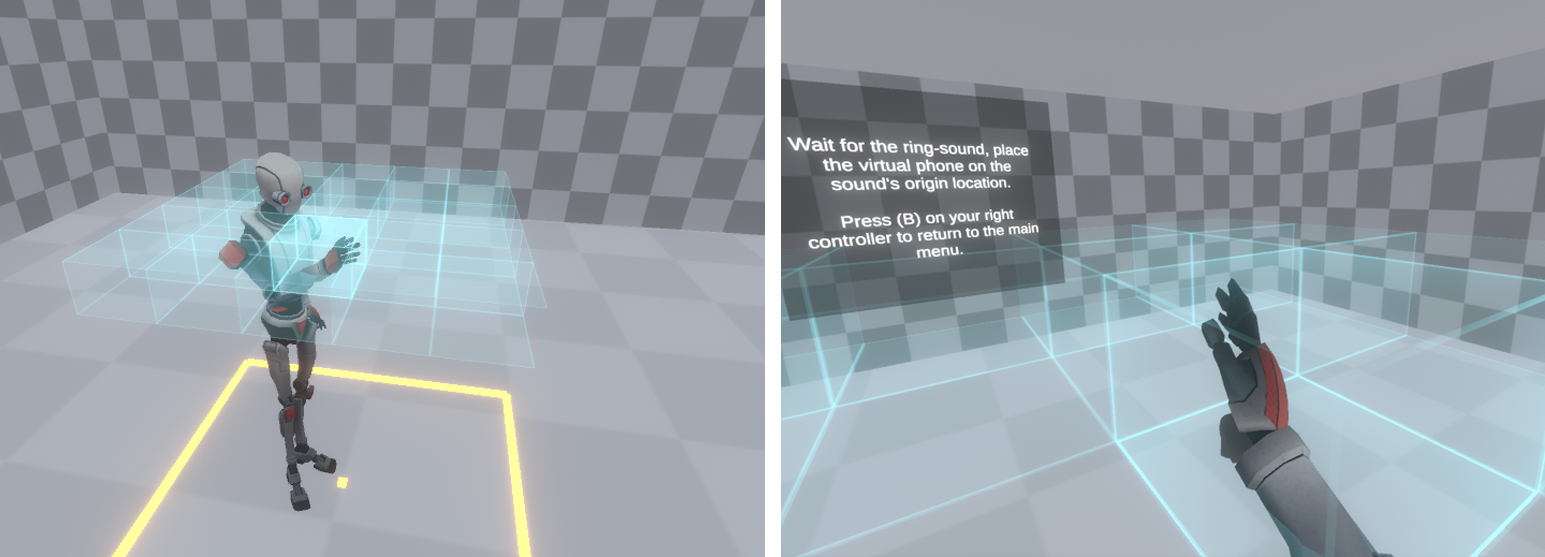}
    \caption{The participant in virtual space, represented by robot, pointing out the sound's location as a highlighted, volumetric cuboid (left) and from first-person view (right)}
    \label{fig:enter-label}
\end{figure}

Observations:

\begin{itemize}
    \item Choosing an appropriately sized volume was difficult and would impact the selection process.
    \item This variant could be obstructive to the participant's view, similarly to the canvas of the laser-pointer version.
    \item No selection outside of head-height or WFS-space was possible.
\end{itemize}

\subsection{WFS-related Problems and Limitations}

Throughout our prototyping stage we encountered a number of challenges that hindered a seamless integration of WFS into a VR application. We list some observed phenomena below relating directly to the concept of our sample experiment.

\subsubsection{Lack of Height Localization}

The underlying WFS setup could only produce wavefronts on a plane at the height of the speaker-array. Participants were informed about this limitation but were nevertheless allowed to place their guesses above or below this height in the versions that allowed it.

\subsubsection{Lack of Occlusion}

The WFS renderer was agnostic to any obstacles located in the virtual environment that a participant might expect to dampen or obfuscate sounds. For instance, ducking behind the bar-counter did not affect sound clarity at all. Room acoustics in general do not conform to those the virtual space would entail and may require an additional step of sound-processing.

\begin{figure}[h]
    \centering
    \includegraphics[scale=0.33]{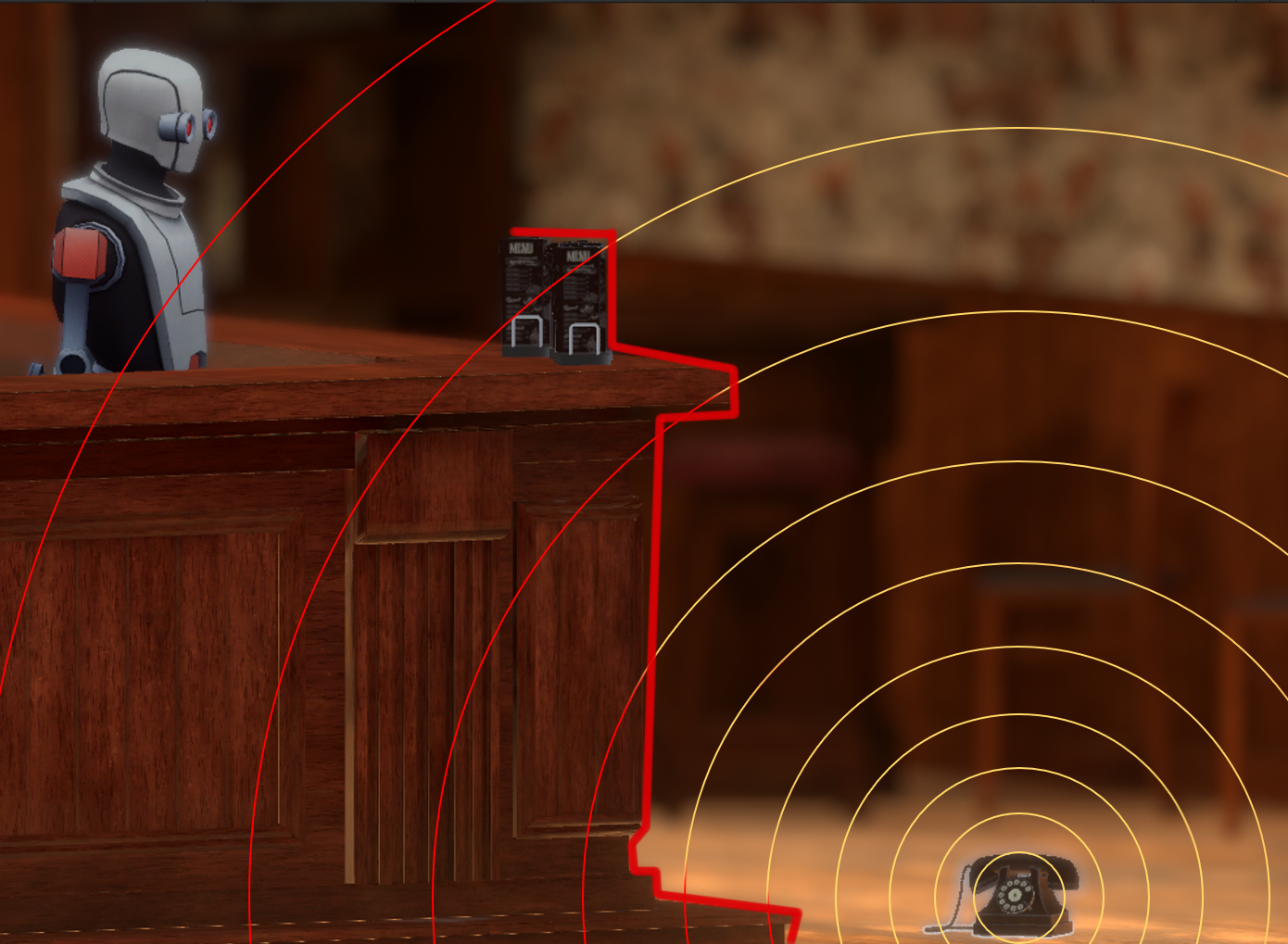}
    \caption{Hypothetical example of a sound being dampened by an occluder. Obstacles surface echoes are not simulated by the WFS renderer.}
    \label{fig:sound-occ}
\end{figure}

As such, the used experiment was only adequate for sound-sources that are in the direct line-of-view of the participant and that were not be expected to create echoes or indirect sound-waves. 

For our final version we chose to employ different types of virtual environments and no additional processing on our sounds. Our results showed no meaningful change in performance when the room's acoustics or aesthetics matched the underlying sound (see fig \ref{fig:env-sound-synergy}).

\subsubsection{Limited Space}

\begin{figure}[h!]
    \centering
    \includegraphics[scale=0.6]{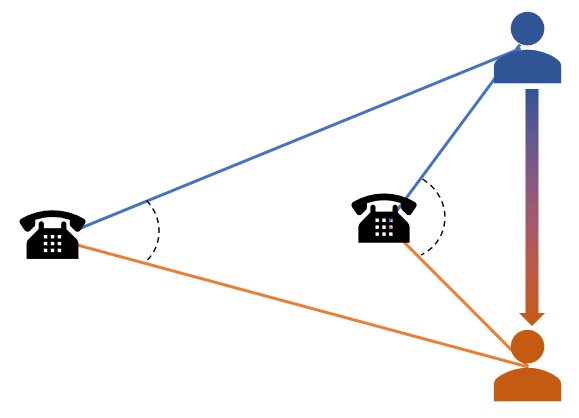}
    \caption{Benefits of moving around the virtual space: The change in direction of a distant phone is lesser than that of a closer one, giving an indication of depth/distance.}
    \label{fig:sound-parallax}
\end{figure}

For purposes of depth perception, moving around the virtual space proved highly beneficial in locating a sound's origin, due to the auditory parallax-effect that would take place whilst moving (see fig. \ref{fig:sound-parallax}).

The employed WFS setup only provided a small area for movement. Participants were instructed that they could move around if they wanted, but not actively encouraged to do so. The virtual environment would display a yellow barrier outlining the bounds of the area they could move in.

\subsubsection{Virtual-Space and WFS-space Alignment}
\label{calib-challenge}

Since the WFS renderer and VR application operated separately, there were some challenges in lining up the visual virtual space of the VR headset with the audio virtual-space of the WFS system.

In addition, the mobile headset and Air Link connection meant that the virtual environment was not placed identically each session. We solved this problem by introducing a manual calibration step before each session using the Quest Pro's pass-through feature to align the environments. See section \ref{spat_calib} for further details.

\subsubsection{Stereo Falloff Parameters}
\label{falloff-challenge}

Because the propagation of sounds largely depends on the acoustics of the surrounding environment, attenuation of virtual sounds is often left to relatively arbitrarily selected parameters that do not necessarily match what it would sound like in reality.

In our case, we chose a \textit{logarithmic} volume rolloff with a minimum distance of $0.1m$ and a maximum distance of $650m$, as well as a spread of $10$ and a neutral Doppler level. Our 3D sound rendering (stereo) was done using Unity's default functionality which does not employ HRTFs.

These settings ensured that the sounds were audible at all distances with a reasonable volume, while still having a clear falloff at larger distances. It was, however, evident that the attenuation was significantly stronger and steeper than that of the WFS system. As such, our data indicated that sounds were easier to locate by attenuation in the stereo condition (see section \ref{search-patterns-ch}).

\subsubsection{Lack of User-Dependent Optimization}
\label{user-dep-opt}

\begin{figure}[h!]
    \centering
    \includegraphics[scale=0.7]{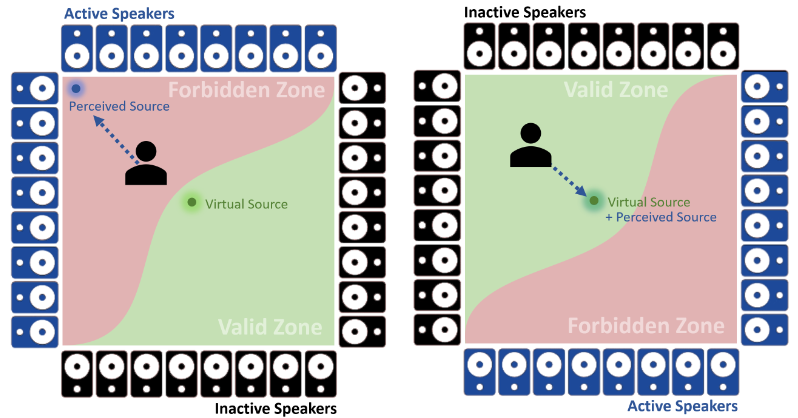}
    \caption{Example of User-Dependent Optimization: The virtual audio source in the center of the listening area is simulated by two speaker arrays, generating a limited area in which the simulation is valid. If the user stands outside this area (left), the user may confuse the sound's origin with that of the active speakers. The WFS renderer can account for this by using a different set of speaker arrays to ensure the user's position is not inside a forbidden zone (right). This requires the system to have real-time data on the user's position.}
    \label{fig:user-adapt}
\end{figure}

The sound-rendering capabilities of WFS are particularly effective for sources positioned outside the array of loudspeakers, which can be precisely reconstructed across the listening field. However, a significant limitation arises when virtual sound sources are to be perceived inside the listening area \cite{Melchior2008UserDependent}. In such scenarios, the accurate reconstruction of the sound field is often only valid for specific, limited regions within the overall listening space (see fig. \ref{fig:user-adapt}) \cite{verheijenPHD}. This means that a listener moving away from these optimal regions may experience shifts in perceived source location. This phenomenon is rooted in the fundamental physical principles and practical implementation constraints of WFS, where perfect reconstruction across an entire volume remains a challenge \cite{verheijenPHD}.

These limitations are directly addressed by Melchior et al. (2008) by performing real-time adaptations of the WFS system based on the user’s tracked position \cite{Melchior2008UserDependent}. This optimization encompasses three key aspects: dynamic secondary source (loudspeaker) selection, real-time delay correction, and adaptive amplitude (level) adjustments. By tailoring the sound field to the listener's instantaneous location, the system aims to minimize the "forbidden zones" where sound reproduction is suboptimal and to ensure consistent sound localization, arrival time, and perceived loudness.

In our study, however, the employed WFS system did not incorporate the user-dependent optimizations described by Melchior et al. The system operates entirely without accounting for the user's position within the listening area. Consequently, we expect the sound localization accuracy to be suboptimal, particularly for sounds located near the center of the listening field. Purely by chance, this inherent limitation might lead participants to gravitate towards specific locations, potentially the edges of the listening area (see section \ref{location-patterns-analysis}).

Other studies involving the localization of WFS sounds in VR have demonstrated a good accuracy \cite{wfs_and_vr}.
\section{Final Version}
\label{final-version-ch}

Based on observations and feedback extracted from the above listed prototypes, we built a version that was ultimately used in data collection and analysis. The version introduced some additional complexity and more conditions in the hopes of yielding more meaningful results.

\subsection{Overview}

Building on insights from previous prototypes, the final version of the experiment was largely based on the \textit{Phone-Placement} prototype (\ref{phone-placer-proto}), but brought along some additional enhancements.

\begin{figure}[h!]
    \centering
    \includegraphics[scale=0.18]{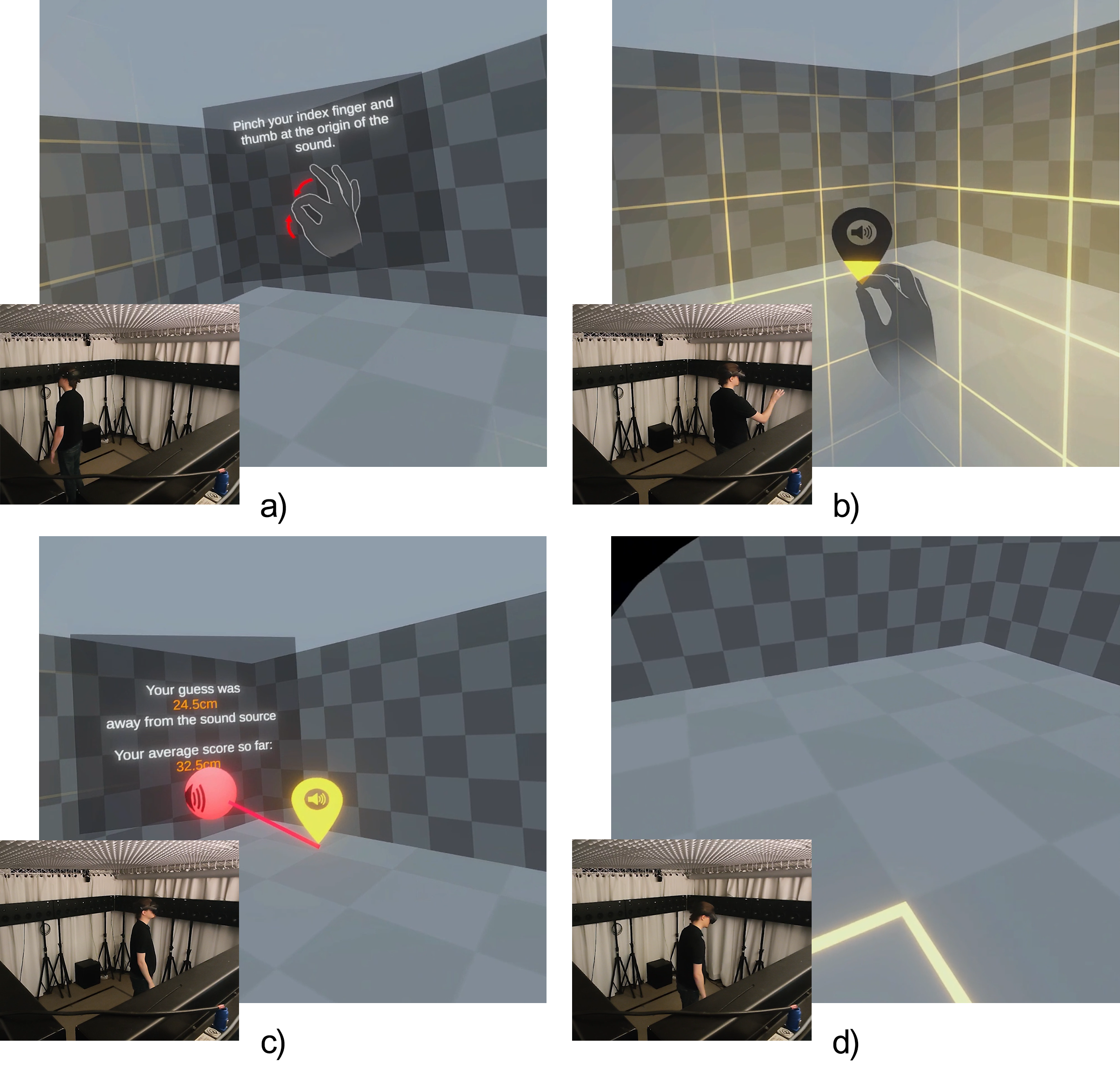}
    \caption{A single trial in the project's final version: The participant hears a sound (a), moves towards it and places a guess on it's origin by pinching their fingers for one second (b). The participant receives feedback by being shown the sound's real location (c). After a while, the indicators disappear and the next trial begins. Videos of full sessions can be viewed under the following links: \url{https://youtu.be/YAaiDir7U2g}, and \url{https://youtu.be/vh2HWklqGWQ}}
    \label{fig:trial_procedure}
\end{figure}

In this version, participants would still be asked to identify the perceived origin of sounds played within a \(2\times2\) m sound-rendering area at head height, but perform this task in varying surroundings and with varying sounds. Participants would indicate their guess by performing a \textit{pinch} gesture at the perceived location using hand tracking. In cases where hand tracking was unreliable, regular VR controllers could be used as a fallback input method.

\subsubsection{Experimental Conditions}

To examine the impact of different auditory and environmental factors, we selected three distinct sound stimuli, each publicly available and varying in frequency and duration:

\begin{itemize}
    \item \textbf{A rotary telephone ring} ($6.12$ seconds long, around 4 of which were highly audible)
    \item \textbf{A short piano melody} ($6.861$ seconds long)
    \item \textbf{A chirping bird in an outdoor environment} ($2$ minutes and $39.362$ seconds long)
\end{itemize}

Participants would only hear the respective sound once per trial and get no repeat attempts. A localization guess could be placed before the sound finished playing.

\begin{figure}[h!]
    \centering
    \includegraphics[scale=0.26]{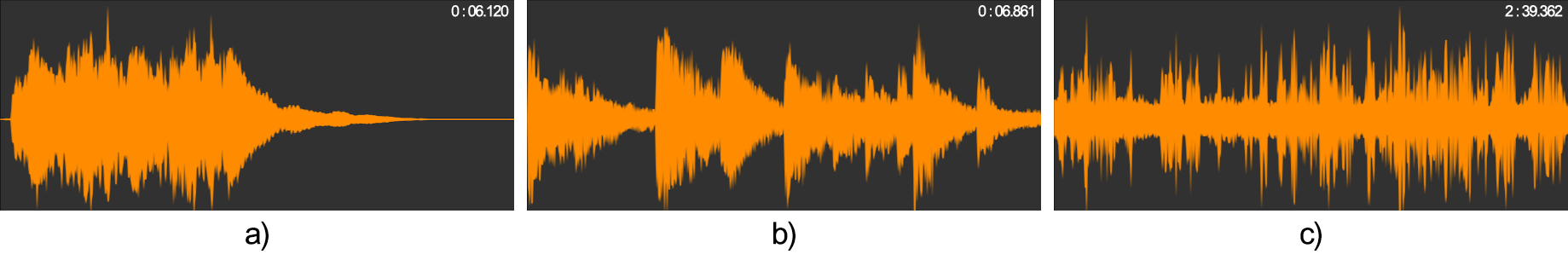}
    \caption{Waveform graphs and duration of the three different sounds used. With a) being the rotary telephone ring, b) the piano tune and c) the birdsong. The clearly audible part of a) lasted four seconds, b) six seconds, while c) lasted for over two minutes, thus usually ending prematurely when the participant placed a guess.}
    \label{fig:sounds}
\end{figure}

To explore whether the \textit{virtual environment} itself influenced localization performance, three different virtual spaces were created to roughly match the nature of these sounds:

\begin{itemize}
    \item \textbf{A neutral, featureless room with checkered walls} (corresponding to the telephone ring)
    \item \textbf{An indoor bar environment} (corresponding to the piano melody)
    \item \textbf{A vast outdoor mountain village} (corresponding to the bird chirp)
\end{itemize}

\begin{figure}[h!]
    \centering
    \includegraphics[scale=0.15]{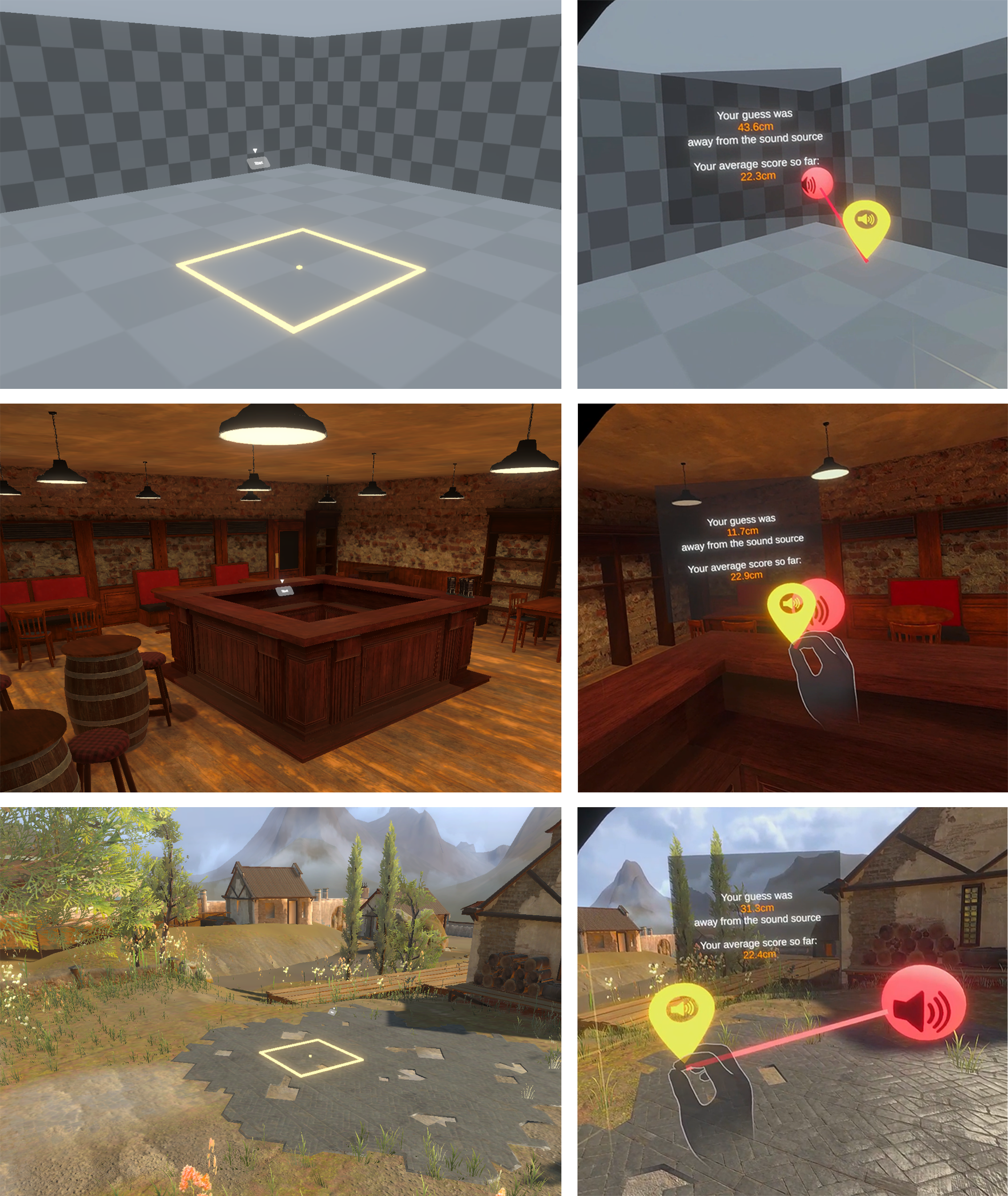}
    \caption{The three virtual environments that participants solved trials in as seen in the Unity Editor (left) and from a participant's perspective after placing a guess (right). The tutorial and dynamic trials were always conducted in the \textit{blank} environment (top).}
    \label{fig:envs}
\end{figure}

Each participant completed \textit{six} trials in each of the three environments (two trials for each sound) in a randomized order. The sound location was chosen at random and would not move throughout the trial.

After completing these 18 \textit{static trials}, participants performed an additional 9 \textit{dynamic trials} (three for each sound), where the sound source moved along a \(2\)-meter trajectory over \(1\)--\(3\) seconds before stopping. These dynamic trials always took place in the neutral, blank environment to isolate movement-related effects.

The entire procedure was repeated for both WFS playback and stereo headphone playback, leading to the following permutations of conditions:

\begin{align*}
    &[\text{WFS} | \text{HMD}] \times [\text{blank} | \text{indoors} | \text{outdoors}] \times 2 \times [\text{sound}_1 | \text{sound}_2 | \text{sound}_3] = 36 \quad (\text{static trials})\\
    &[\text{WFS} | \text{HMD}] \times 3 \times [\text{sound}_1 | \text{sound}_2 | \text{sound}_3] = 18 \quad (\text{dynamic trials})
\end{align*}

This resulted in a total of 54 trials per participant, with an average completion time of approximately 25-35 minutes.

\begin{figure}[h!]
    \centering
    \includegraphics[scale=0.3]{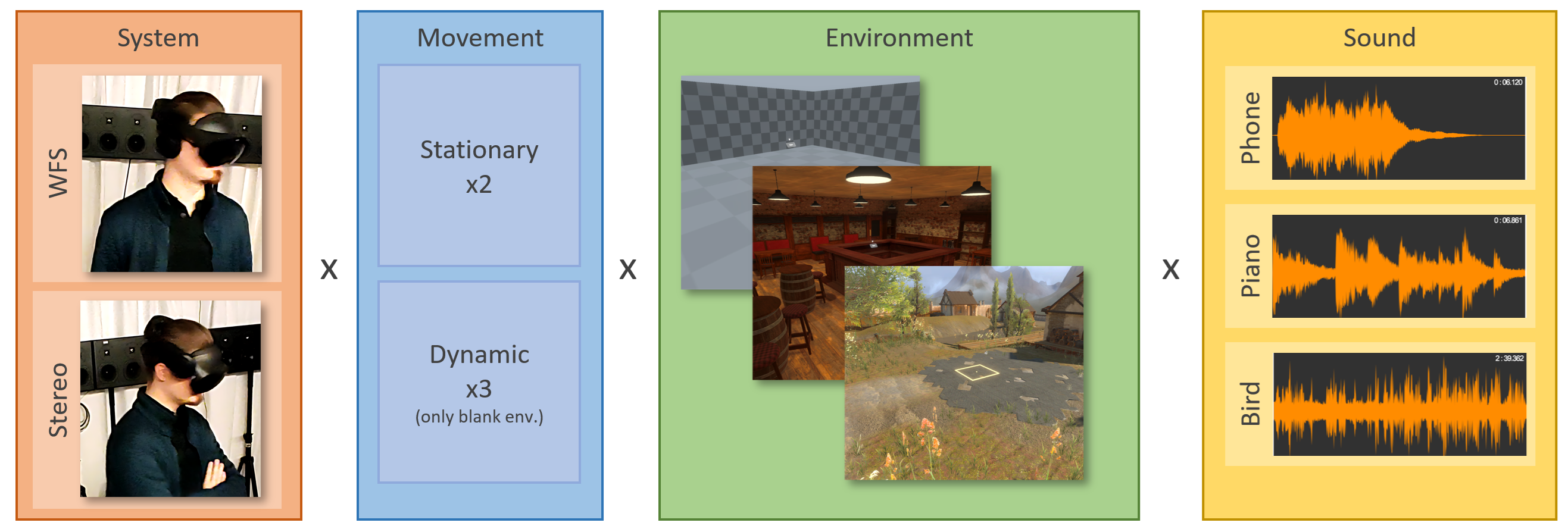}
    \caption{Overview of the individual parameters, producing a total of 24 unique permutations (conditions). Each stationary condition would be performed twice, and each dynamic one thrice by each participant. Trials would be randomized within each block.}
\end{figure}

\subsubsection{Research Questions}

Beyond evaluating the accuracy of WFS vs. stereo playback, the experiment was designed to explore additional research questions, including:

\begin{itemize}
    \item \textbf{Impact of the virtual environment}: Does the visual environment affect sound localization performance?
    \item \textbf{Impact of sound type \& duration}: Do different sound types and durations influence accuracy?
    \item \textbf{Environment-sound synergy}: Does performance improve when a sound matches its expected environment (e.g., a piano melody in a bar)?
    \item \textbf{Moving sound sources}: How does localization accuracy change when sound sources are moving, and do these effects differ between WFS and stereo?
\end{itemize}

By systematically varying these factors, the experiment aimed to provide a comprehensive assessment of the viability of WFS for VR-based auditory research.

\subsection{Tutorial}
\label{tutorial}

Before starting the experiment, participants were guided through an interactive tutorial designed to familiarize them with the trial procedure and the method for placing guesses.

The tutorial was narrated using AI-generated voice lines (produced with ElevenLabs\footnote{\url{https://elevenlabs.io/}}) and provided step-by-step instructions. It introduced participants to the experiment's mechanics, including:

\begin{itemize}
    \item How to indicate the perceived location of a sound using hand tracking (or controllers, if needed).
    \item How trials would be structured, including the distinction between static and moving sounds.
    \item Safety and comfort disclosures, such as showing the virtual barrier participants could not pass through, or explaining that the experiment could be paused or aborted at any time in the case of motion sickness or other reasons.
\end{itemize}

\begin{figure}[h]
    \centering
    \includegraphics[scale=0.13]{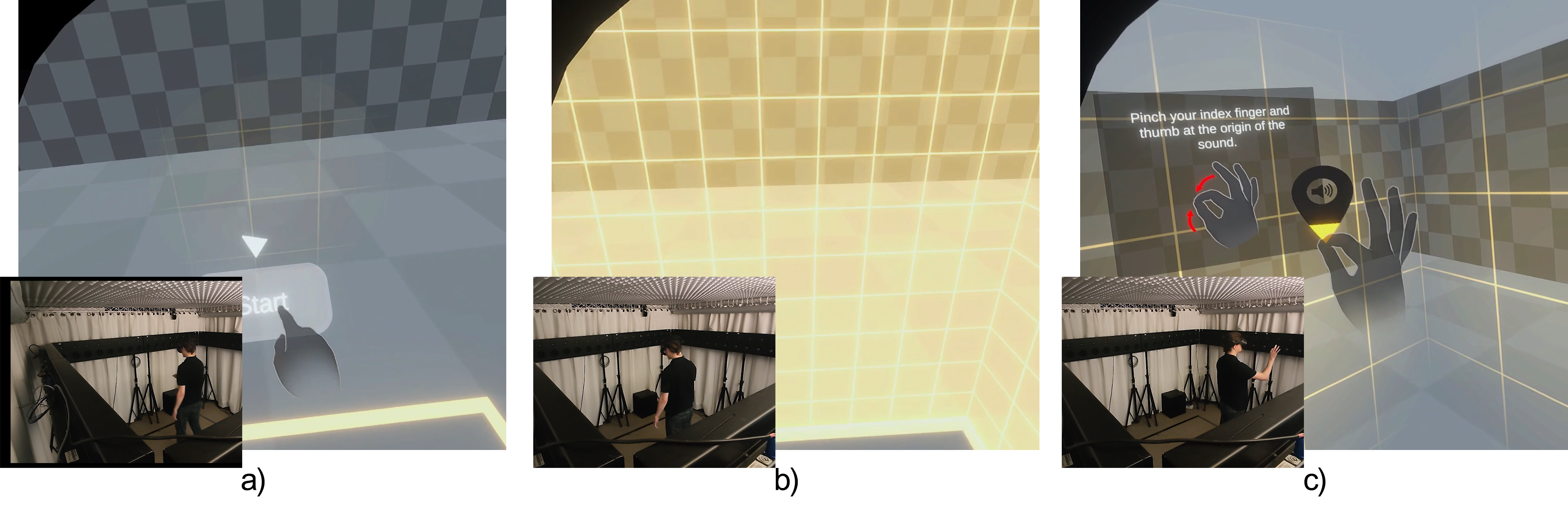}
    \caption{The stages of the interactive tutorial taking place before a session. The session began when the participant pressed the 'Start'-button (a), they were subsequently briefed on the experiment, including the yellow safety-barrier separating the walkable VR space (b). The participant was then guided through four training trials. The tutorial ran through AI-generated voice-lines, a transcript of which can be viewed in section \ref{voicelines-transcript}. A video of the tutorial can be viewed at the beginning of this video: \url{https://youtu.be/YAaiDir7U2g}.}
    \label{fig:tutorial}
\end{figure}

Following the tutorial, participants completed four test trials, featuring one with explicit instructions, and another three for each sound type. These test trials allowed them to get accustomed to the controls and the process of submitting guesses.

The tutorial always took place in the neutral checkered environment using the WFS system to maintain consistency. Additional audio announcements and instructions were played at key moments, such as switching between virtual environments, transitioning between WFS and Stereo playback or beginning the moving-sound trials.

\subsection{Spatial Calibration}
\label{spat_calib}

As described in section \ref{calib-challenge}, a major challenge in using a mobile VR headset was ensuring that the virtual space and the WFS space were perfectly aligned. Since small discrepancies could significantly impact the accuracy of spatial sound localization, we introduced a calibration step that was performed by the experimenter before each session, prior to the participant putting on the headset.

\begin{figure}[h!]
    \centering
    \includegraphics[scale=0.2]{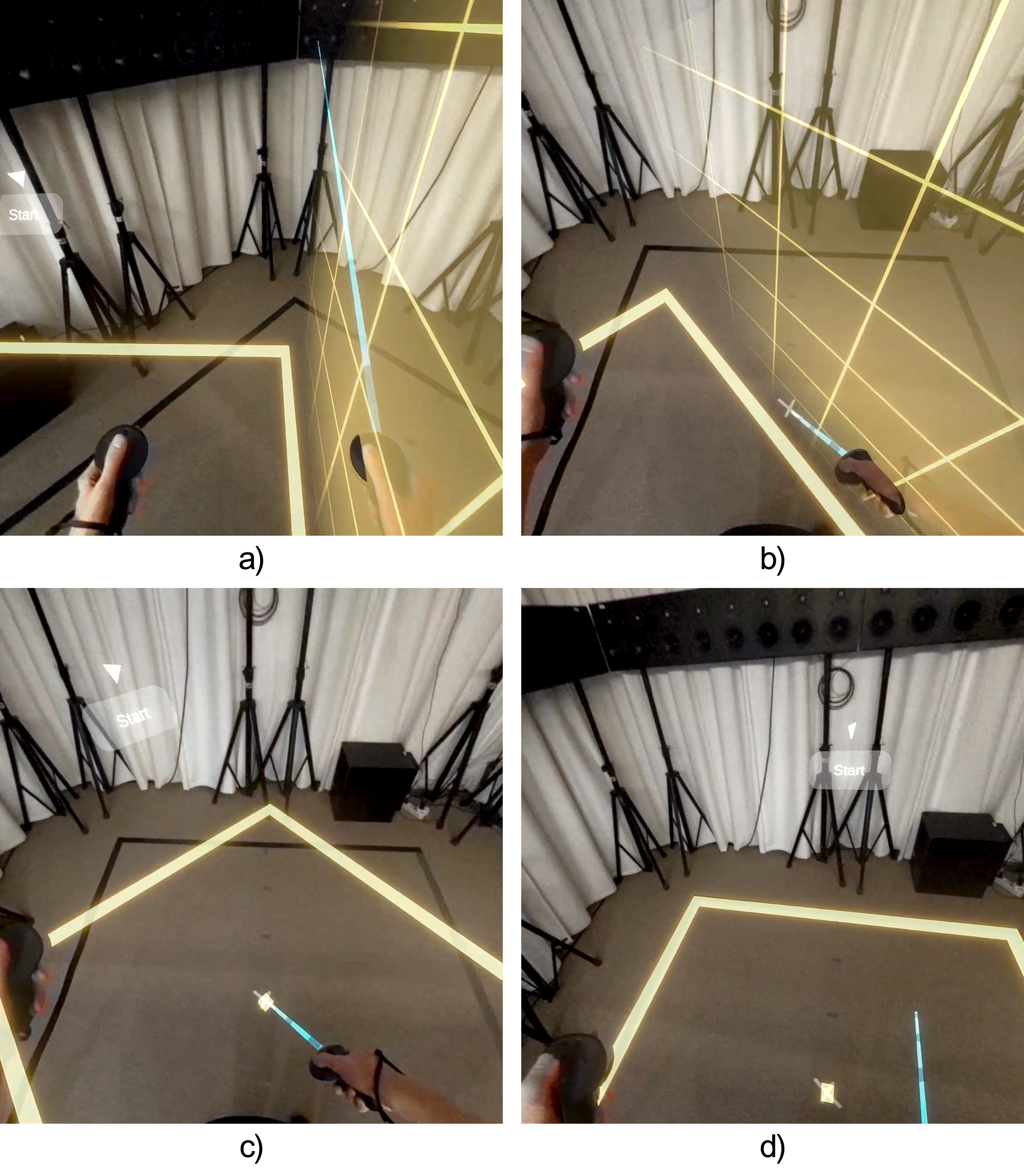}
    \caption{VR-view of the calibration steps performed before every session. The calibration mode showed the passthrough-image of the real world, overlayed with the yellow highlights delineating the virtual space. At first, the two spaces are missaligned (a). The first step consists of pointing the controller at the center-spot of the real wfs-space and pressing the trigger button (b); this lines up the centers. The virtual wfs-space is then rotated around the center using the analogue stick on the left controller (c), leading to the two spaces being fully aligned (d). A video of the calibration procedure can be viewed here: \url{https://youtu.be/668u1TINfV0}}
    \label{fig:calibration_steps}
\end{figure}

To streamline this process, our Unity program leveraged the Quest Pro’s passthrough functionality. At the experiment’s starting screen, a specific button combination on the VR controllers would activate calibration mode, temporarily removing the virtual environment and replacing it with the passthrough image of the real world. Over this image, a virtual overlay displayed the boundary and center of the WFS area.

Using the VR controllers, the experimenter could manually align the virtual and real-world WFS zones. The right controller functioned as a laser pointer, allowing the experimenter to indicate the “true” center of the WFS space. Pressing the trigger button would then reposition the virtual zone accordingly. To finalize the alignment, the experimenter could use the left controller’s analog stick to rotate the virtual space around this center point until its boundaries matched those of the real WFS area.

This process ensured that the spatial mapping of sounds in VR accurately corresponded to their real-world WFS locations, although we expect that some degree of error remained. The alignment error would only be present in WFS conditions, as stereo playback didn't rely on an external system.

\subsection{Experimenter Controls}

Because we used an Air Link connection, the experimenter was able to continuously monitor the participant’s view through a mirrored display on the intermediary PC. In addition to this mirrored view, the experiment interface also displayed the current status of the experiment, including which trial was in progress, the sound being played, and the participant’s responses.

\begin{figure}[h]
    \centering
    \includegraphics[scale=0.35]{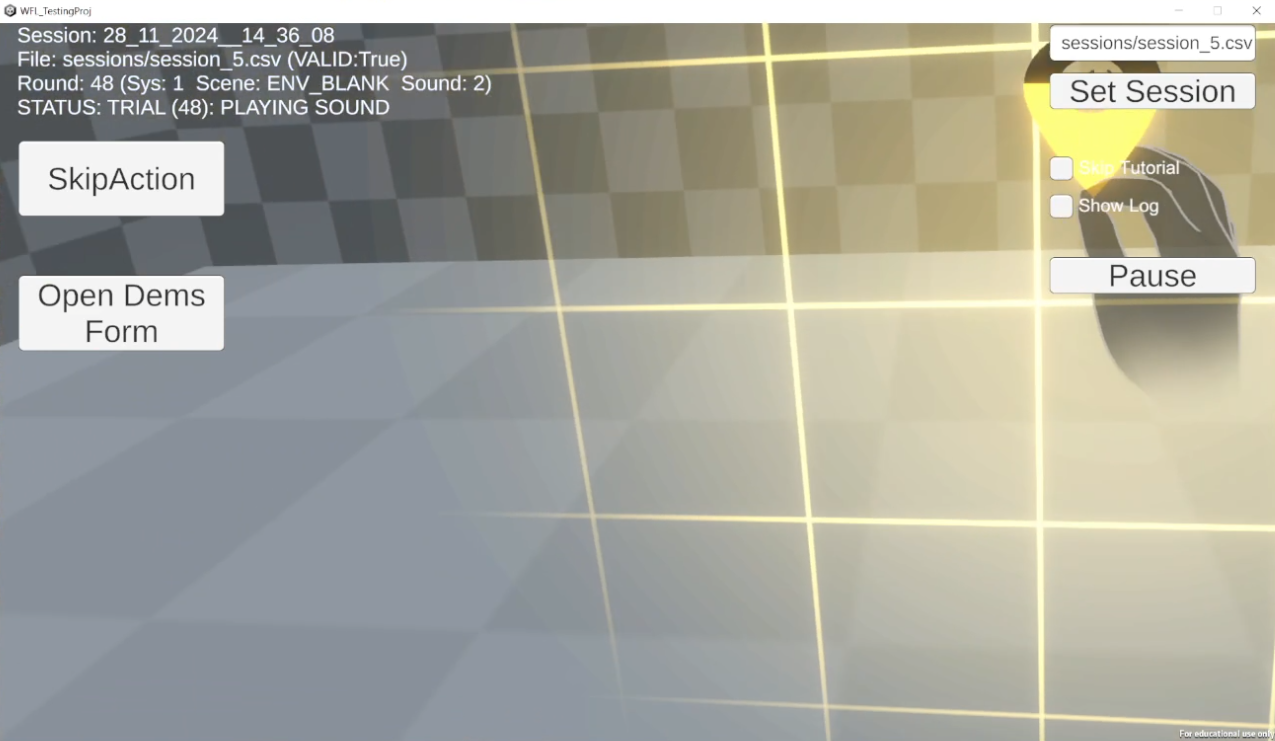}
    \caption{Screenshot of the application the experimenter could see during a session. The application shows a mirrored view of the participant's left eye, in addition to the current status and some controls.}
    \label{fig:experimenter-controls}
\end{figure}

To maintain control over the session, the experimenter had the ability to manually skip steps or pause the experiment if necessary. The system also included an automatic pause function that would activate if the participant removed the headset, ensuring that no trials continued without their awareness.

At the end of each session, participants were prompted to enter demographic information through a simple form. This form could be accessed and toggled directly from the experiment window, allowing for a smooth transition from the experimental phase to data collection.

\begin{figure}[h]
    \centering
    \includegraphics[scale=0.435]{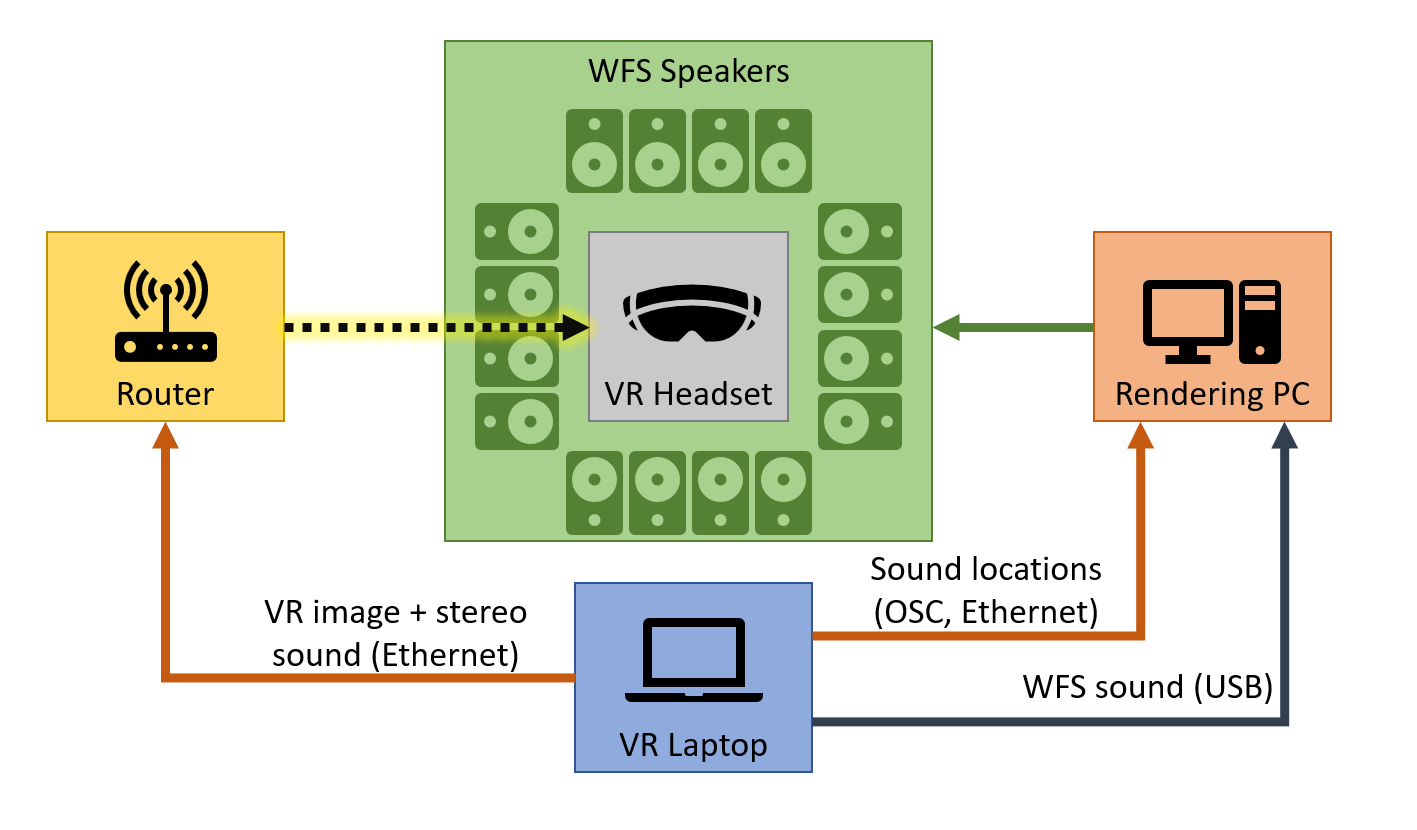}
    \caption{Hardware setup of our experiment. An intermediary VR laptop would render the VR image and send it to the HMD via a router while also managing sound playback of the WFS system via the WFS render PC.}
    \label{fig:hardware-setup}
\end{figure}

\subsection{Participants and Data Collection}

The experiment was conducted with six participants (four male, two female) aged between 27 and 40. All participants were recruited internally at the MPIB as volunteers. Most of them reported having either a "regular" or "enthusiast" level of prior VR experience.

Due to technical difficulties with switching between audio devices, the majority of participants (four out of six) began the experiment using the WFS system before transitioning to the stereo condition in the second half of the session.

During each session, we recorded the positions and timestamps of both the played sounds and the participants' guessed locations. Additionally, continuous tracking data was collected for the position and rotation of the head-mounted display and both hands. Since the Quest Pro relies on its built-in cameras for hand tracking, reliable tracking data was not always available. In cases where tracking was lost, default values were assigned, which were later filtered out by our analysis scripts.

The results of data collection are examined and analyzed in the next chapter.

\begin{figure}[h]
    \centering
    \includegraphics[scale=0.2]{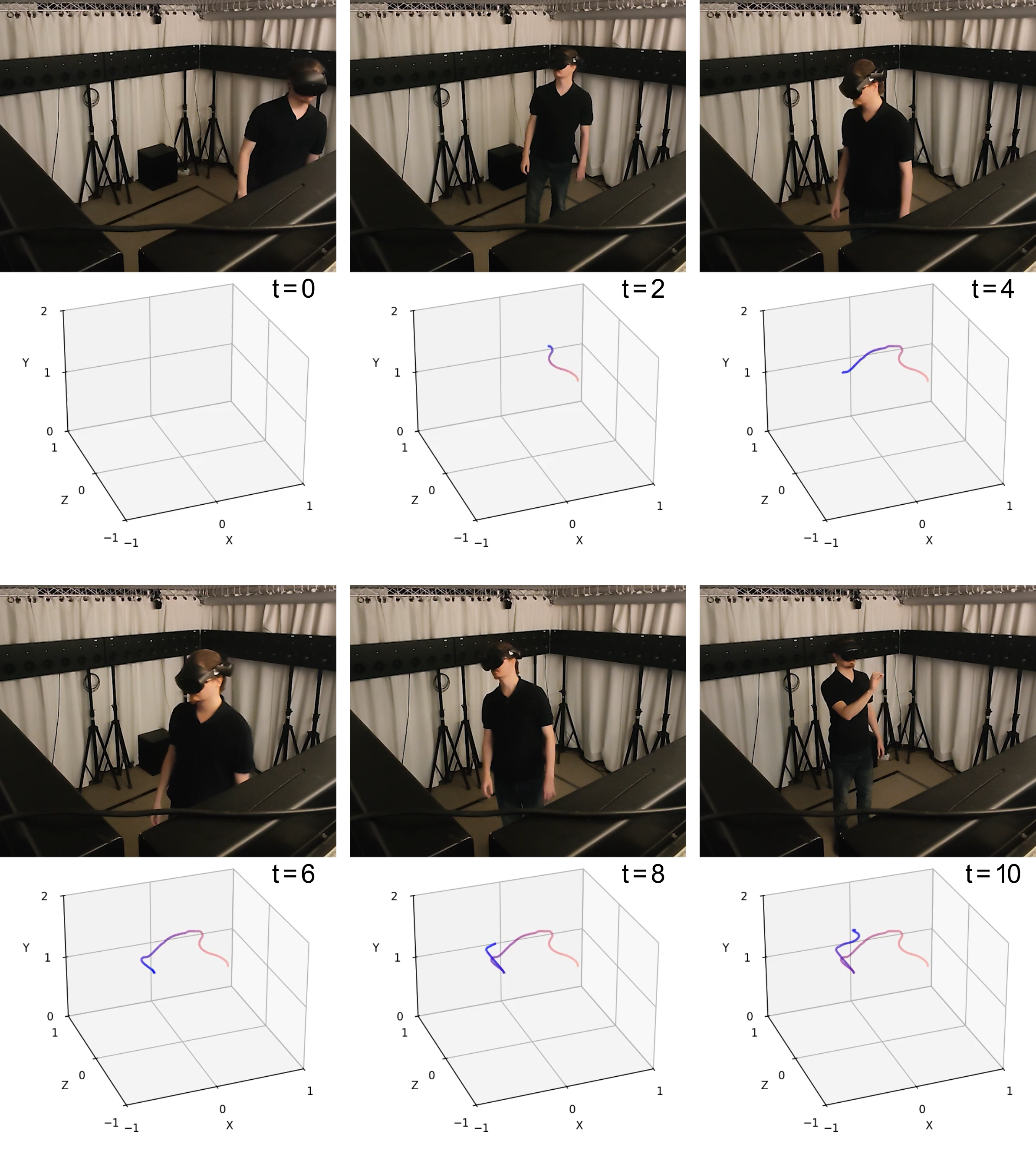}
    \caption{The tracked headset path for subsequent timestamps (in seconds) of a single trial. At $t=2$ the participants hears a sound and tries to assess it's location before placing a guess at $t=10$. A video of this trial (including tracking data) can be viewed here: \url{https://youtu.be/lkfHRkv3YhI}}
    \label{fig:tracking}
\end{figure}
\section{Analysis}
\label{analysis}

\subsection{Localization Accuracy}

\begin{figure}[h]
    \centering
    \includegraphics[scale=0.36]{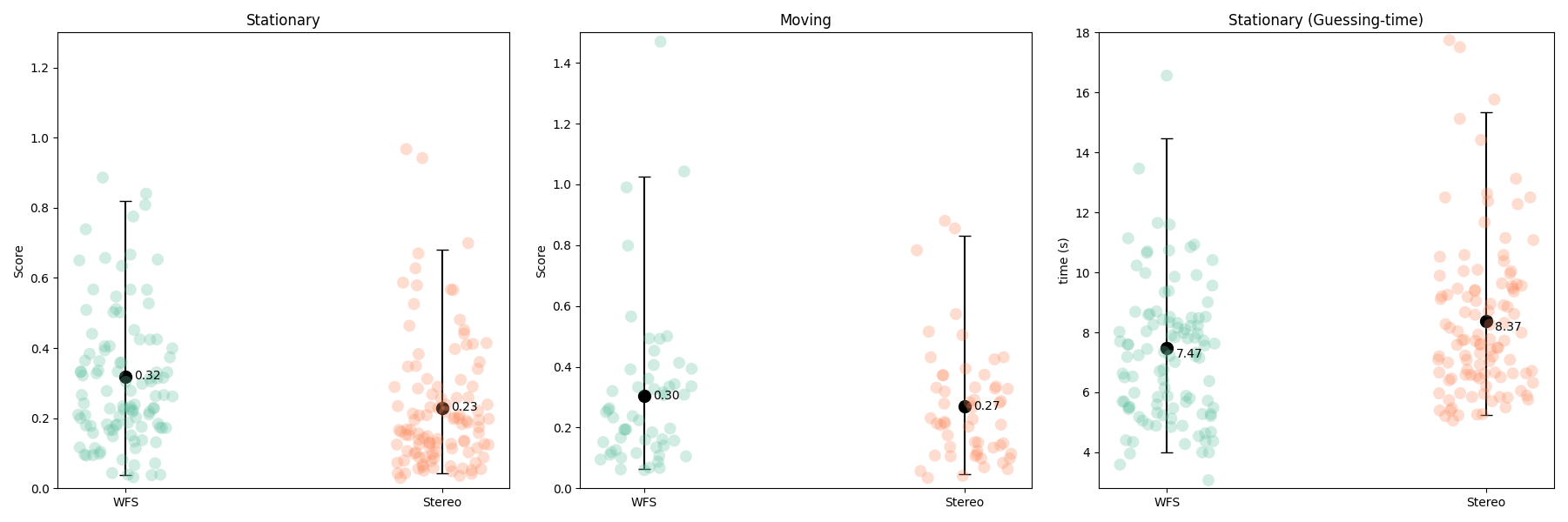}
    \caption{Average score (guess-to-sound distance, lower is better) across all trials for the stationary condition (left), dynamic condition (middle), and the average time to place a guess in stationary trials (right). Participants took longer in placing a sound in stereo conditions, but tended to be more accurate.}
    \label{fig:performance-general}
\end{figure}

Across all 324 trials (54 per 6 participants), results indicate that participants took longer in localizing sounds using stereo, but that the additional search-time paid off in consistently achieving a higher accuracy when compared to the WFS system. In particular, guesses with an error margin of less than 20 cm nearly doubled in the stereo condition relative to WFS. However, as shown in fig. \ref{fig:performance-general}, this discrepancy was somewhat reduced when participants localized moving sounds.

Despite the lower accuracy scores, many participants reported that the WFS system produced a more natural auditory experience and provided a stronger initial sense of direction. It is important to consider that the 3D sound rendering settings used in the experiment were configured with a steep distance-based falloff and did not induce calibration-based errors (see section \ref{falloff-challenge} and \ref{spat_calib}), providing a strong localization advantage to stereo trials. Meanwhile, the lack of \textit{user-dependent optimization} (see section \ref{user-dep-opt}) acts as an inherent disadvantage for the WFS system which can be corrected \cite{Melchior2008UserDependent}.

\begin{figure}[h]
    \centering
    \includegraphics[scale=0.36]{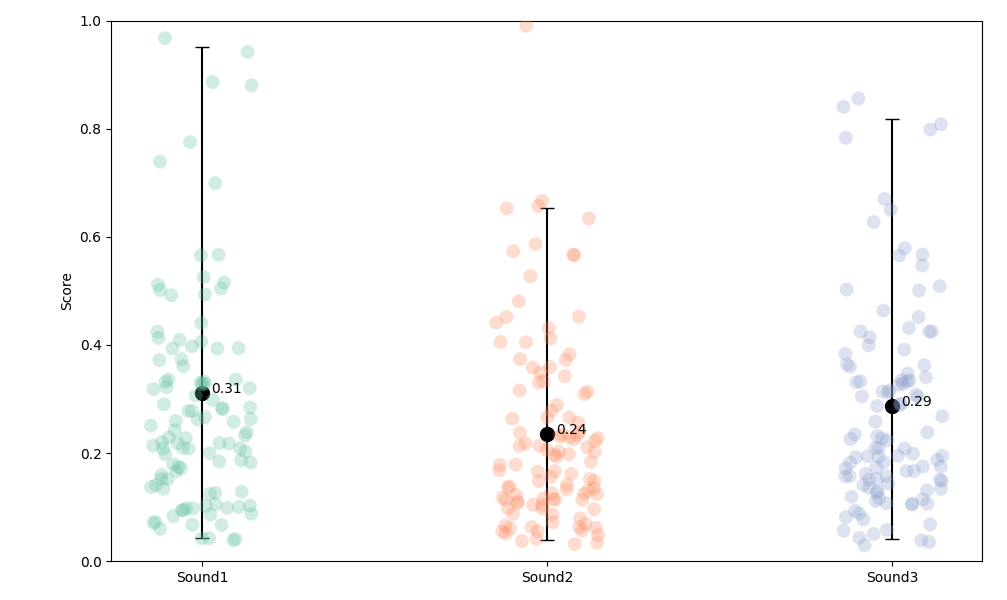}
    \caption{Average scores (lower is better) for each of the three sounds used. Participants had the best score with the longest sound and the worst score with the shortest one.}
    \label{fig:sound-type-perf}
\end{figure}

Performance also varied by sound type. As shown in fig. \ref{fig:sound-type-perf}, participants achieved the highest accuracy with Sound 2 (birdsong) and the lowest with Sound 1 (telephone), with performance directly correlating to the length of each sound. These differences persist throughout the varying conditions.

\begin{figure}[h]
    \centering
    \includegraphics[scale=0.43]{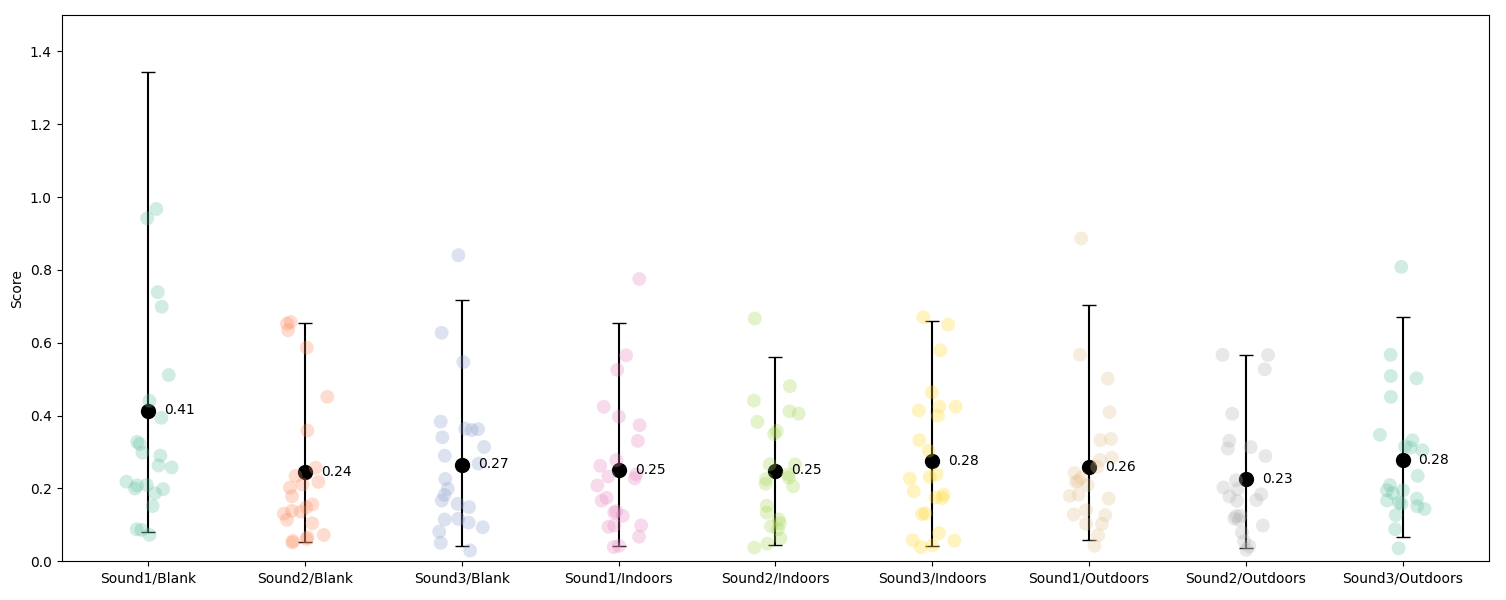}
    \caption{Average scores (lower is better) for each combination of virtual environment and sound. There was no conclusive effect derived from thematically matching sounds/environments.}
    \label{fig:env-sound-synergy}
\end{figure}

When examining performance across the three virtual environments (see fig. \ref{fig:env-sound-synergy}), accuracy remained relatively consistent, with the exception of a slight decrease in the blank (neutral) environment. A closer look at the interaction between sound type and environment suggests that certain sounds \textit{may} have been easier to localize when presented in a thematically matching environment. However, the small sample size of these highly specific conditions prevents us from drawing any definitive conclusions about the impact of virtual environments on localization performance.

\subsection{Learning Effects}

\begin{figure}[h]
    \centering
    \includegraphics[scale=0.38]{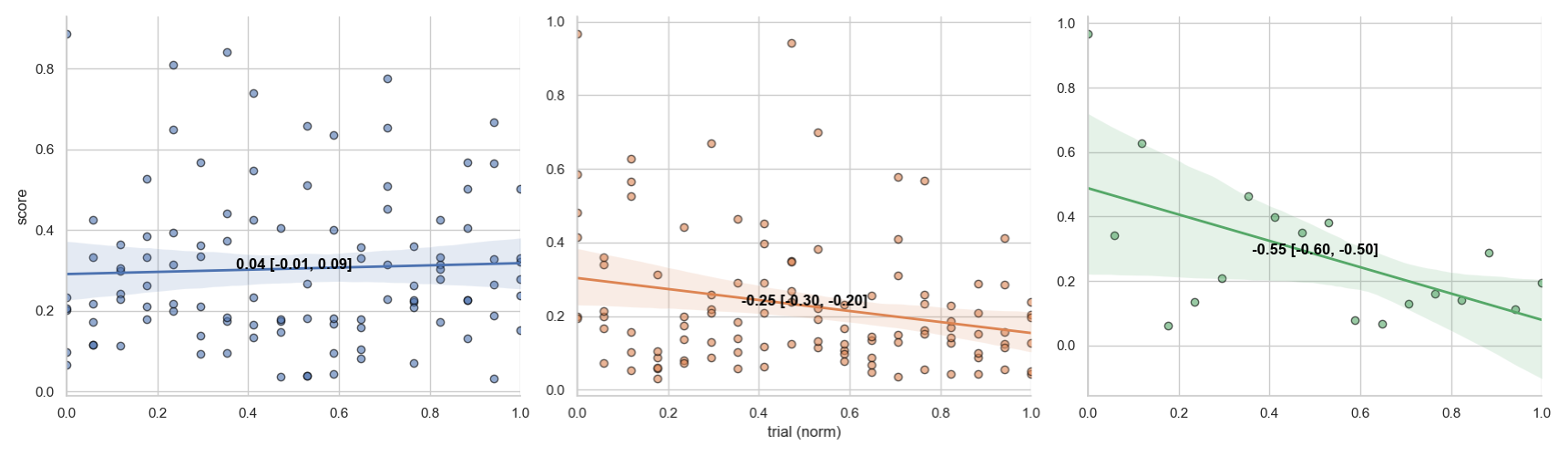}
    \caption{Linear regression of all scores (lower is better) in stationary trials, ordered chronologically. Participants demonstrated little to no improvement in performance with the WFS system (left), but did so with stereo (middle). The participant with the highest level of improvement (right), achieved so using the stereo system.}
    \label{fig:learning-rates}
\end{figure}

Participants exhibited a greater learning rate in the stereo condition compared to the WFS condition, as reflected in the overall trend of improving scores over time. A linear regression shown in fig. \ref{fig:learning-rates} of accuracy across trials confirms this pattern.

Of the six participants, five showed a positive learning rate in the stereo condition, with a regression slope of less than -0.1, indicating steady improvement. In contrast, only one participant demonstrated a significant learning trend in the WFS condition, suggesting that performance in WFS remained relatively stable across trials.

This pattern supports the idea that WFS provides a more intuitive localization experience, where participants can rely on their natural spatial hearing without needing extensive adaptation. Stereo, on the other hand, offers higher potential accuracy but requires participants to adjust to its limitations—particularly the steep volume attenuation, which provides less immediate directional accuracy but enhances precision once participants engage in movement-based localization strategies.

\subsection{Location Patterns}
\label{location-patterns-analysis}

Sound positions were randomly generated within the walkable area, resulting in an isotropic and generally even distribution across trials. Even so, participants demonstrated a distinct preference for placing their guesses near the edges and corners (particularly in the upper-right corner) when using the WFS system. On the other hand, participants were likelier to place their guesses in a central location when using the stereo headphones.

This pattern suggests that WFS sound sources originating near the center of the space were more difficult to localize precisely. While the general direction was clear, exact placement proved more challenging. This aligns with our expectations outlined in section \ref{user-dep-opt}, supporting the fact that WFS setups \textit{without} user-dependent optimization stochastically shift perceived sound sources towards the edges of the listening area.

Based on this observation, we expect such a WFS system to out-perform a stereo system for trials in which sound-sources are located \textit{outside} the listening area, such as the \textit{Dot-Circle prototype} described in section \ref{dot-circle-proto}.

\begin{figure}[h]
    \centering
    \includegraphics[scale=0.43]{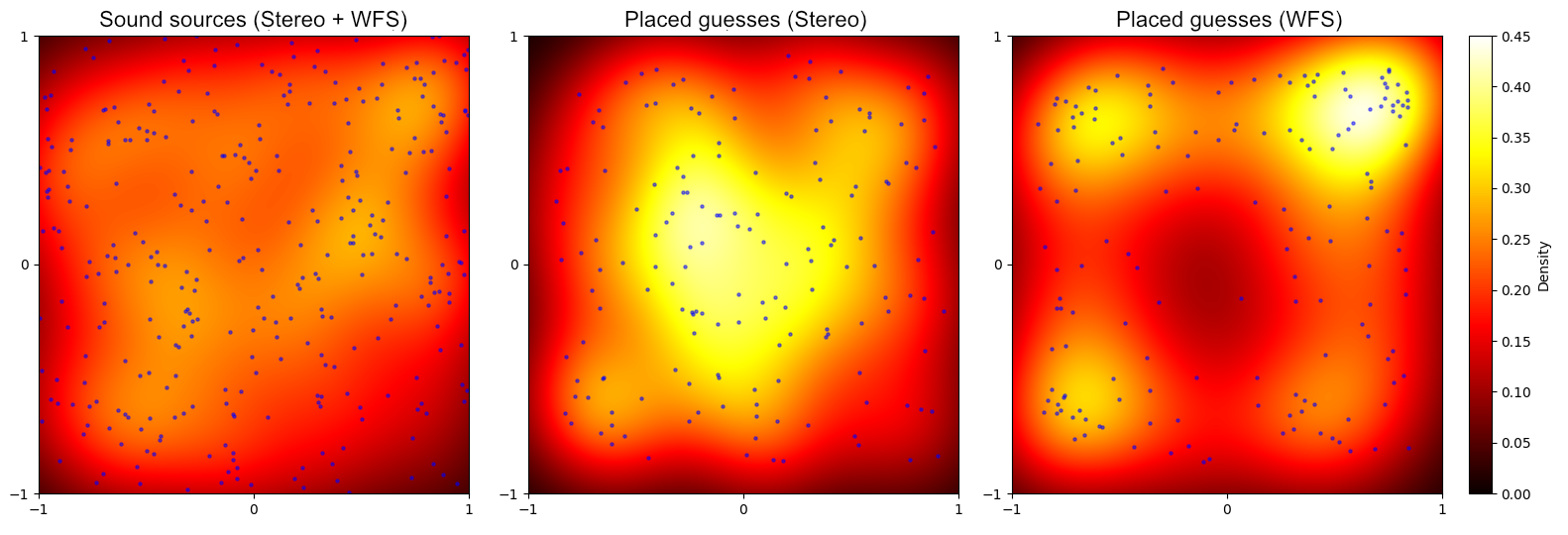}
    \caption{Heatmap showing density of points for \textit{all} generated sound-sources (left), guesses placed in stereo conditions (middle) and guesses placed in WFS conditions (right). Participants consistently tended to place their guesses towards the corners and edges in WFS, whilst sticking closer to the center in stereo.}
    \label{fig:heatmaps}
\end{figure}

\begin{figure}[h]
    \centering
    \includegraphics[scale=0.45]{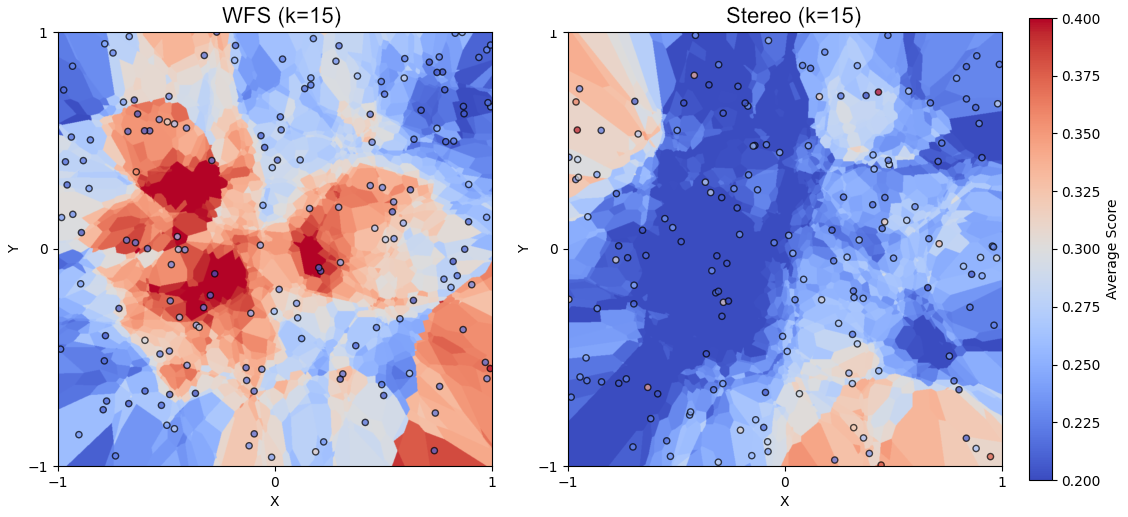}
    \caption{Weighted kNN ($k=15$) graph of the average score (lower is better) across the playable area. Participants tended to perform better in WFS when the sound-source was near the edge, while doing the opposite in stereo.}
    \label{fig:knn_map}
\end{figure}

Correspondingly, as shown in fig \ref{fig:knn_map}, accuracy in WFS was highest when the sound source was near the speaker arrays, whereas in stereo trials, performance was more evenly distributed, with a slight preference toward central placements.

This supports the idea that WFS provided strong directional cues but struggled with precise localization when the sound originated near the center. Conversely, stereo allowed for finer-grained assessment of the sound’s position, as participants could move closer and rely on volume differences to pinpoint the source more accurately.

This high preference towards the edges in WFS implies that WFS systems are more adequate for use-cases in which sounds originate from outside the playable area and where the direction of the origin is more important than it's distance.

\subsection{Search and Movement Patterns}
\label{search-patterns-ch}

The average trial followed a predictable procedure for all participants: the distance between the right hand (used for placing guesses) and the sound source steadily decreased, reaching its lowest point toward the end of the trial. The movement of the HMD followed a similar trajectory, except that its distance from the sound source tended to increase slightly at the very end. Participants were also highly likely to be looking directly at the sound source when placing their final guess, as opposed to at the start of the trial.

\begin{figure}[h]
    \centering
    \includegraphics[scale=0.5]{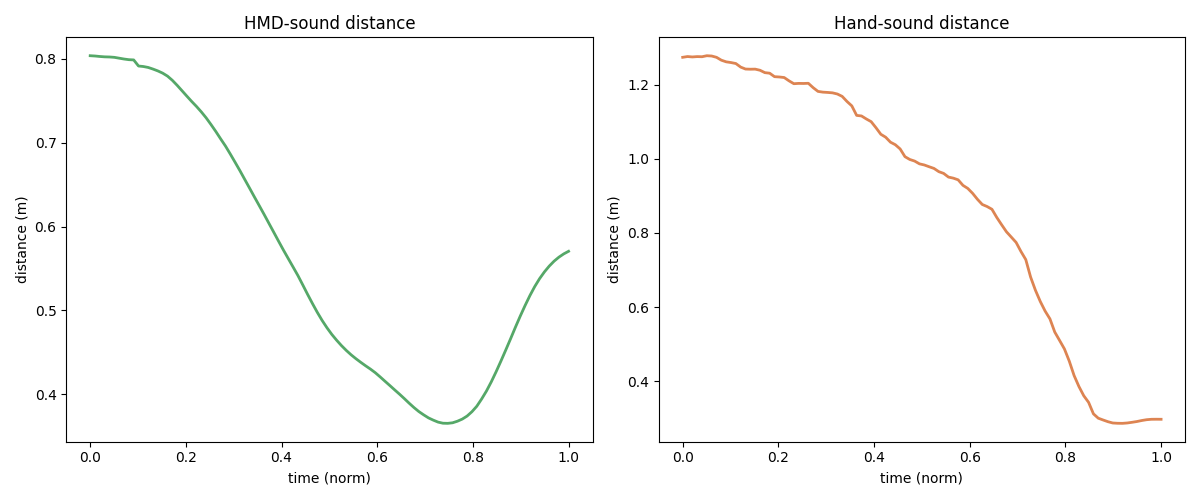}
    \caption{Average distance of HMD to the sound-source (left) and the participant's right hand to the sound-source (right) throughout all trials, between hearing a sound and placing their guess (normalized time). Participants would use their head to locate the sound source, then lean back and place their guess using their hand.}
\end{figure}

Participants primarily relied on their head movement to scan for the sound source before becoming confident enough to place their guess. Once they identified the location, they tended to lean back slightly, swapping their head with one of their hands to place a guess.

\subsubsection{Search Paths}

While individual movement patterns varied significantly, making broad generalizations difficult, certain trends consistently emerged based on the sound condition.

\begin{figure}
    \centering
    \includegraphics[scale=0.4]{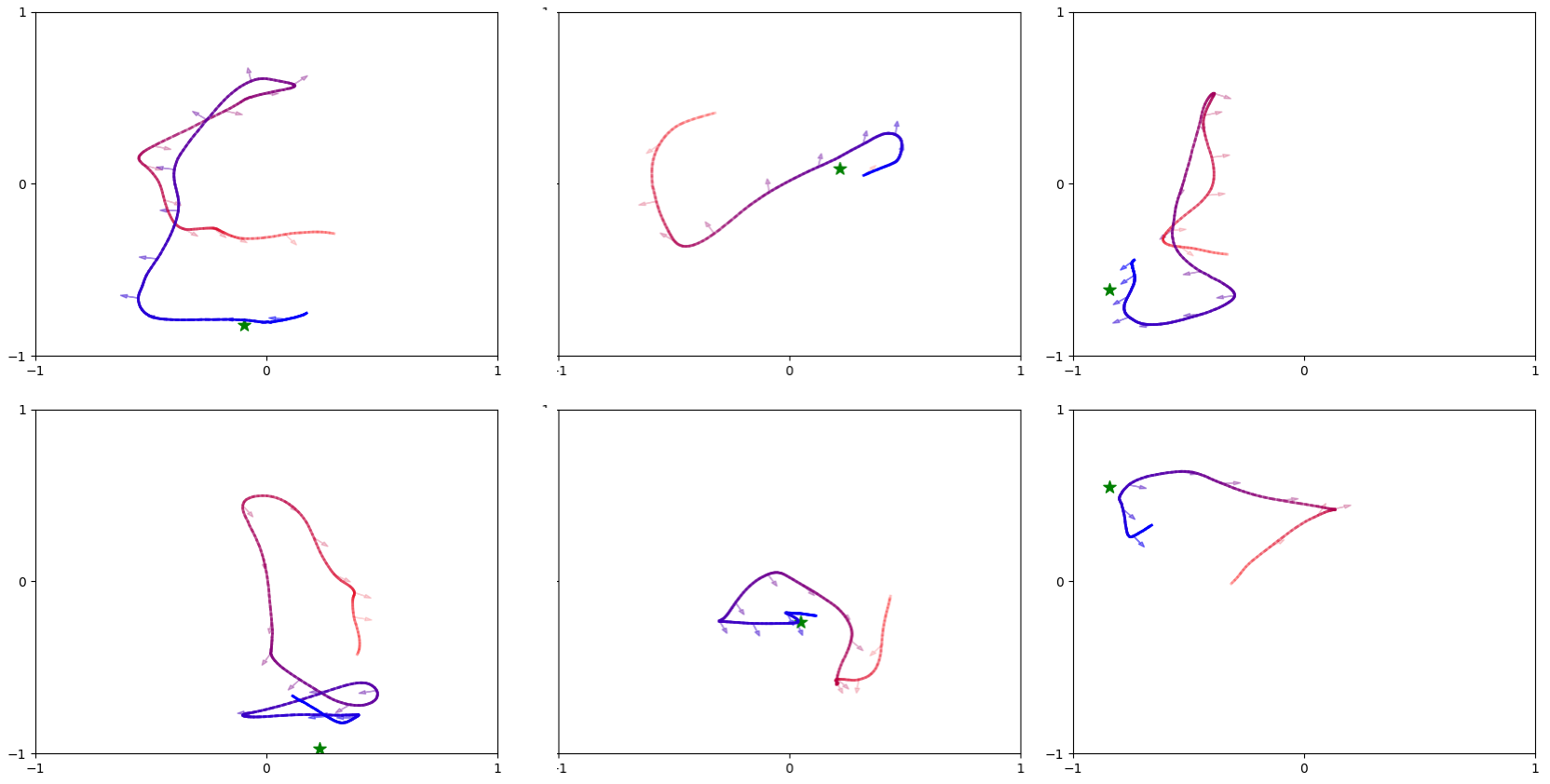}
    \caption{Examples of a common search patterns observed in stereo rounds: The participant would often move away from the sound-source initially, then quickly adjust their trajectory based on volume attenuation. Their trajectory towards the sound source would often be less direct than in WFS.}
    \label{fig:stereo-paths}
\end{figure}

\begin{figure}
    \centering
    \includegraphics[scale=0.4]{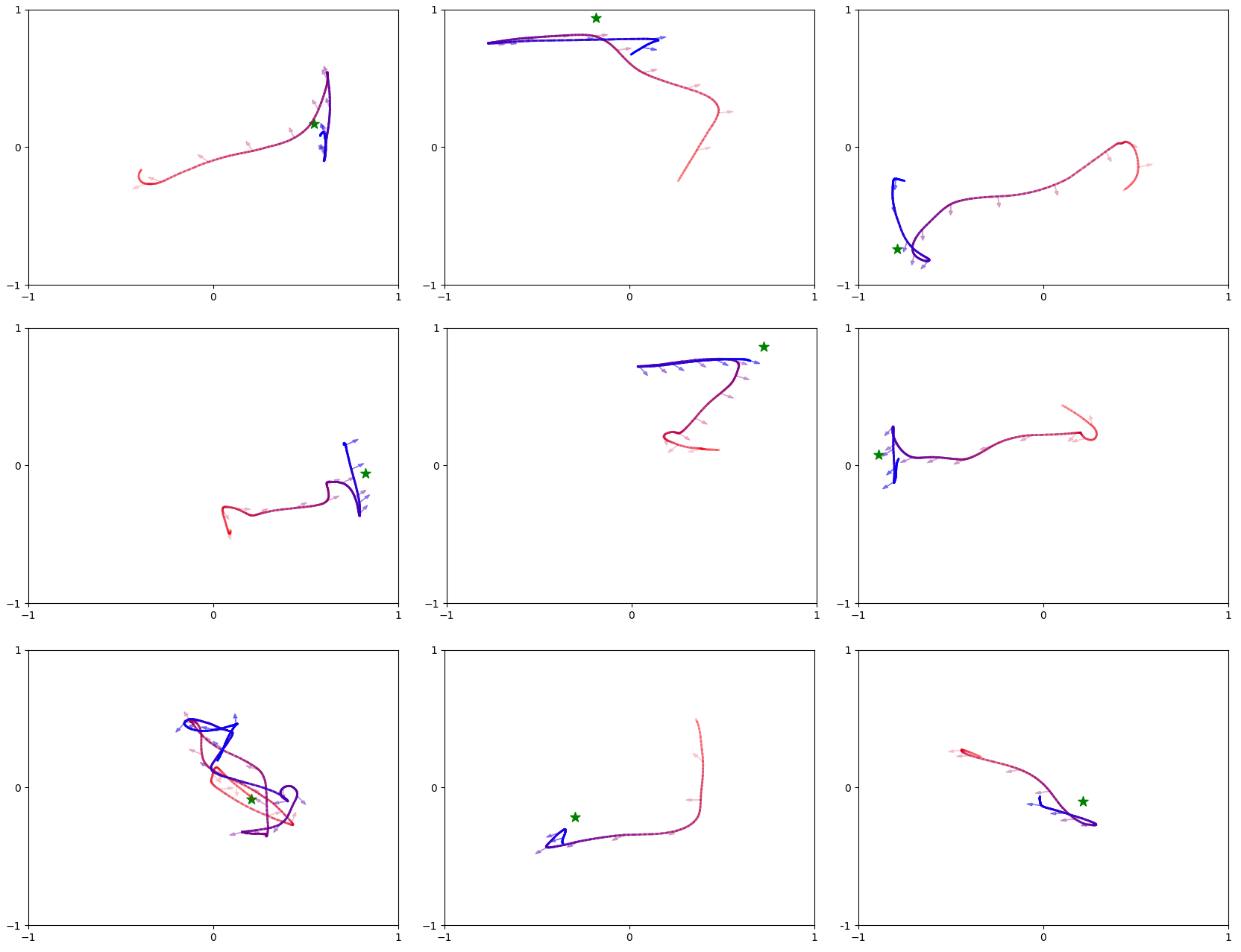}
    \caption{Examples of a common pattern exhibited in WFS trials: The participant would approach the correct direction immediately, then refine their search by moving in parallel to the speaker array. For sound sources located near the center (bottom row) movements were less consistent.}
    \label{fig:wfs-paths}
\end{figure}

As can be observed in fig. \ref{fig:stereo-paths}, the most common strategy in successful stereo trials involved maintaining a moderate distance from the sound source for most of the trial. Instead of moving directly toward it, participants often moved away at first, likely using volume attenuation to establish a directional reference. After a series of adjustments along different axes, the sound’s position would become apparent, leading to a relatively confident final guess.

WFS trials with high accuracy (see fig. \ref{fig:wfs-paths}) followed a different pattern. Participants typically identified a rough direction almost immediately by detecting the active speakers within the array. However, instead of moving directly toward the sound, they frequently spent extended periods shifting parallel to the speaker array, refining their estimate by comparing subtle differences in the perceived wavefronts.

For WFS trials where the sound source was positioned near the center of the room, movement patterns became noticeably more erratic. These trials rarely produced high scores, as participants struggled to identify a clear reference point. The lack of distinct speaker cues in the middle of the space led to hesitation and a less structured search process, reinforcing the idea that WFS systems require user-dependent optimization (see section \ref{user-dep-opt}) to render sounds inside the listening area with a reliable accuracy.

\section{Conclusions}

The use of WFS in VR-based cognitive and behavioral studies presents both unique challenges and distinct advantages.

While its implementation demands greater effort from developers, it offers a more natural and intuitive interface for participants, potentially enhancing ecological validity. When paired with a mobile headset and hand-tracking controls, WFS minimizes the need for participants to learn complex controls, such as button mappings or specific interaction techniques. Instead, they can rely on their natural body movements and sense of hearing, making the experience more accessible.

As the technology matures, we anticipate a reduction in development effort, but remain skeptical that the additional cost and endeavor is worthwhile for studies where the sense of hearing is secondary. A standardized API for sound positioning and playback would greatly simplify implementation, eliminating the need for OSC-based location setting and separate interfaces for sound playback.

From both the development process and data analysis, we identified several key observations:

\begin{itemize}
    \item A stereo setup is more flexible, particularly for simulating effects like occlusion. While localization accuracy in stereo can surpass WFS, it requires adaptation to artificial attenuation patterns.
    \item Participants in stereo trials rely more on perceiving volume attenuation, whereas in WFS, they depend more on their natural binaural hearing.
    \item The accuracy gap between stereo and WFS is smaller for moving sound sources than for static ones.
    \item WFS, by using speakers and artificial wavefronts, provides a more natural-sounding soundscape and an intuitive interface. WFS is a clear winner in use-cases where direction matters more than exact location or distance, where as regular headphones/HRTFs are more adequate for use-cases where participants walks close and among the sound emitters.
    \item WFS systems without \textit{user-dependent optimization} (see section \ref{user-dep-opt}) are less accurate than stereo systems when the sounds are \textit{inside} the listening area due to physical limitations. We expect WFS localization to improve substantially by incorporating the user's position into the sound rendering process.
    \item Longer-duration sounds are easier to localize.
    \item The visual setting of the virtual environment has no measurable impact on performance.
\end{itemize}

This study was limited to isolated point-source sounds. We expect WFS to be more suitable for research involving complex soundscapes with multiple directions and sources, such as those studied in soundscape psychology and psychoacoustics. Other experiments utilizing WFS and VR have demonstrated satisfactory localization accuracies \cite{wfs_and_vr}.

While we look forward to further advancements in this field, we anticipate that WFS will remain a specialized tool for research focused on auditory perception, rather than a mainstream addition for enhancing immersion in general VR studies.

\section{Literature}

\resetboldbibnames
\newrefcontext
\DeclareNameAlias{author}{FirstLast}
\printbibliography[heading=none,notkeyword=own,notkeyword=ownpreprint,resetnumbers=true]

@article{Bosman02012024,
author = {Isak de Villiers Bosman, Oğuz ‘Oz’ Buruk, Kristine Jørgensen and Juho Hamari},
title = {The effect of audio on the experience in virtual reality: a scoping review},
journal = {Behaviour \& Information Technology},
volume = {43},
number = {1},
pages = {165--199},
year = {2024},
publisher = {Taylor \& Francis},
doi = {10.1080/0144929X.2022.2158371},
URL = {https://doi.org/10.1080/0144929X.2022.2158371},
eprint = {https://doi.org/10.1080/0144929X.2022.2158371}
}

@article{KaiyanZhu,
author = {Zhu, Kaiyan and Lin, Rong and Li, Hong},
year = {2021},
month = {12},
pages = {},
title = {Study of virtual reality for mild cognitive impairment: A bibliometric analysis using CiteSpace},
volume = {9},
journal = {International Journal of Nursing Sciences},
doi = {10.1016/j.ijnss.2021.12.007}
}

@article{Faria2023,
  author    = {Ana L. Faria and Jorge Latorre and M. Silva Cameirão and Sergi Bermúdez i Badia and Roser Llorens},
  title     = {Ecologically valid virtual reality-based technologies for assessment and rehabilitation of acquired brain injury: a systematic review},
  journal   = {Frontiers in Psychology},
  volume    = {14},
  pages     = {1233346},
  year      = {2023},
  month     = {August 29},
  doi       = {10.3389/fpsyg.2023.1233346},
  pmid      = {37711328},
  pmcid     = {PMC10497882}
}

@article{Parsons2015,
  author    = {Thomas D. Parsons},
  title     = {Virtual Reality for Enhanced Ecological Validity and Experimental Control in the Clinical, Affective and Social Neurosciences},
  journal   = {Frontiers in Human Neuroscience},
  volume    = {9},
  pages     = {660},
  year      = {2015},
  month     = {December 11},
  doi       = {10.3389/fnhum.2015.00660},
  pmid      = {26696869},
  pmcid     = {PMC4675850}
}

@INPROCEEDINGS{WFS_og,
  author={Brandenburg, Karlheinz and Brix, Sandra and Sporer, Thomas},
  booktitle={2009 3DTV Conference: The True Vision - Capture, Transmission and Display of 3D Video}, 
  title={Wave Field Synthesis}, 
  year={2009},
  volume={},
  number={},
  pages={1-4},
  doi={10.1109/3DTV.2009.5069680}}

@article{Berkhout1993,
  author    = {A. J. Berkhout and D. De Vries and P. Vogel},
  title     = {Acoustic Control by Wave Field Synthesis},
  journal   = {The Journal of the Acoustical Society of America},
  volume    = {93},
  month     = {May},
  year      = {1993},
  pages     = {2764--2778}
}

@article{So2010,
  author    = {Richard H. So and Brian Ngan and Andrew Horner and Jonas Braasch and Jens Blauert and Kwok L. Leung},
  title     = {Toward orthogonal non-individualised head-related transfer functions for forward and backward directional sound: cluster analysis and an experimental study},
  journal   = {Ergonomics},
  volume    = {53},
  number    = {6},
  pages     = {767--781},
  year      = {2010},
  month     = {June},
  doi       = {10.1080/00140131003675117},
  pmid      = {20496243}
}

@phdthesis{verheijenPHD,
    author = {E.N.G. Verheijen},
    title = {Sound reproduction by wave field synthesis},
    school = {TU Delft},
    year = {1998},
    url = {https://resolver.tudelft.nl/uuid:9a35b281-f19d-4f08-bec7-64f6920a3821}
}

@inproceedings{wfs_and_vr,
    author = {Springer, Jan P. and Sladeczek, Christoph and Scheffler, Martin and Hochstrate, Jan and Melchior, Frank and Frohlich, Bernd},
    title = {Combining Wave Field Synthesis and Multi-Viewer Stereo Displays},
    year = {2006},
    isbn = {1424402247},
    publisher = {IEEE Computer Society},
    address = {USA},
    url = {https://doi.org/10.1109/VR.2006.33},
    doi = {10.1109/VR.2006.33},
    booktitle = {Proceedings of the IEEE Conference on Virtual Reality},
    pages = {237–240},
    numpages = {4},
    series = {VR '06}
}

@inproceedings{Melchior2008UserDependent,
  author = {Melchior, Frank and Sladeczek, Christoph and de Vries, Diemer and Fröhlich, Bernd},
  title = {User-dependent optimization of wave field synthesis reproduction for directive sound fields (AES Convention Paper 7376)},
  booktitle = {Presented at the 124th AES Convention},
  year = {2008},
  month = {May},
  address = {Amsterdam, The Netherlands}
}

\newpage
\section{Addendum: Technical Documentation}
\label{tech-documentation}

The complete source code, including the Unity project and analysis framework/scripts can be accessed here: \url{https://arc-git.mpib-berlin.mpg.de/arc-support/unity/wfs_vr}. The project contains licensed assets and is thus only internally accessible to MPIB members.

\subsection{Setup}

Our application ran on a Lenovo Legion 5 Pro that was connected via Ethernet to the private network of the WFS rendering PC, as well as via another Ethernet connection to a 5GHz router placed inside the WFS area that would connect to the Quest Pro headset via an Air Link Wifi connection. In addition, the Lenovo Laptop was connected via a USB connection to the RME MADIface USB with the respective driver installed to play sounds on the WFS system from within the Unity application (For WFS sounds we used a simple \verb|SoundSource| component that was placed directly on top of the \verb|AudioListener| component).

\subsection{Unity Project}

The project uses Unity 2022.3.14f1 with the Universal Render Pipeline. For the best performance, the project was built into an executable using the IL2CPP compiler before being used with participants.

The initial prototype was built using the in-house VR development framework ARC-VR\footnote{\url{https://github.com/MPIB/arc-vr}}. However, as the project matured and the decision was made to integrate hand-tracking and pass-through, large parts of the project were ported over to Meta's own OculusVR framework. As such, the primary camera rig follows the principles of the Oculus framework, while some individual components (like the calibration laser-pointer) remain ARC-VR-based.

The custom scripts written for this project are categorized by 'modules', which can be found in the \verb|Assets/Modules/| directory. Most of these scripts are simple utility scripts or effects. We will provide a basic outline on the more specialized ones below.

\subsubsection{GameMaster}

The \verb|GameMaster| is a general supervisor script that handles almost all of the experiment's behavior, such as running trials, playing sounds, loading or logging data. Most other modules are managed by it.

\subsubsection{SceneLoader}

The core of the entire application is contained in the \verb|MainScene|. This scene contains all of the logic, scripts and camera used throughout the entire game. It also contains the \textit{blank} environment type (see fig. \ref{fig:envs}).

The other environments (\textit{indoors} and \textit{outdoors} scenes) contain \textit{only} the visuals of these scenes and no game logic. The scenes are all loaded \textit{additively} onto the MainScene at the start of the application (via the \verb|SceneLoader| script) and then enabled/disabled when needed. This ensures a seamless transition between scenes, at the cost of a longer startup duration.

\subsubsection{DeviceSwitcher}

Unfortunately, Unity does not support switching audio-devices natively. As such, we use a custom \verb|DeviceSwitcher| module that handles switching audio-devices between WFS and stereo.

It does this by invoking the \textit{svcl-x64}\footnote{\url{https://www.nirsoft.net/utils/sound_volume_command_line.html}} command-line tool with custom parameters to disable and switch audio devices. The component needs to be configured with an appropriate path to the SVCL executable as well as the corresponding names of the virtual devices.

We had some problems in cases where the experiment began with WFS, then switched to Stereo and then back again to WFS. We assume that the problem was caused by Meta's Air Link streaming connection and would not occur in setups without it.

\subsubsection{ElevenLabs Plugin}

To convert text into spoken voice lines for the tutorial and scene transitions, we used ElevenLabs\footnote{\url{https://elevenlabs.io/}}. For ease of development, we used a simple Editor script that used ElevenLabs' HTTP API to automatically generate MP3 files of the provided text and save them into the project.

\subsubsection{OSC API}
\label{docs:osc-api}

We used \textit{extOSC}\footnote{\url{https://github.com/Iam1337/extOSC}} as a general interface to send OSC messages to the WFS render PC. On top of this, we used a custom \verb|OSC_API| script that allowed playing sounds at a certain location or on a certain trajectory through a simple function call.

\subsubsection{Sessions}

The \verb|GameMaster| script automatically sets up a session based on the session filename provided through the experimenter control interface. The session files are contained in \verb|Assets/sessions/| and are pre-generated and randomized using the contained \verb|generate.py| script.

Each line in a session file describes a trial by listing the respective sound, environment, system, and movement. The application handles scene and system transitions automatically.

\subsubsection{Data Logging}
\label{logging}

Data is logged as CSV files into the \verb|Logs/| folder with a name consisting of the date and time that the session was started on. The logged data consists of a \verb|session.csv| file that lists general information on each trial such as the position of a sound, the timestamp of the guess as well as its position.

Each trial also comes with a separate \verb|pos_round_[trialNr].csv| file that contains continuous data on the position and rotation of the headset and the participant's hands (or controllers, if available). This data is logged at a rate of $50Hz$, with timestamps being relative to the entire session.

If hand-tracking was used and the headset could provide no sensible data for a hand, the value would revert to a default value, which we filter out in our Python analysis scripts (see section \ref{analysis-scripts}).

Each session also comes with a \verb|dems.csv| file that contains the data inserted into the demographic data form.

There are many more components to our Unity project. Most of them have intuitive naming or code-comments as on-the-fly documentation.

\subsection{Analysis Scripts}
\label{analysis-scripts}

Our Python framework is contained inside the \verb|Analysis/| directory and is tailored to the data provided by our logging components (see section \ref{logging}).

We load the data of each session into a custom \verb|session| object that contains the metadata from the \verb|session.csv| and \verb|dems.csv| files, as well as a list of custom \verb|trial| objects which contain information and positional data on each trial.

From there, we use a combination of Python's \verb|matplotlib| and various derivatives to form heatmaps, plot positions, compute smooth averages, or kNN maps. Each type of graph used in the analysis section (\ref{analysis}) was created using a separate Python script.

\section{Addendum: Link List}

Project source code: \url{https://arc-git.mpib-berlin.mpg.de/arc-support/unity/wfs_vr/-/tree/master}

Playlist of video demonstrations: \url{https://www.youtube.com/watch?v=CWmqsejxNqI&list=PLh6z44emfoZ3JXKUnx9NbELTxwcd0YAhp&ab_channel=BenjaminKahl}

Initial prototype video: \url{https://youtu.be/CWmqsejxNqI}

Full session videos: \url{https://youtu.be/vh2HWklqGWQ}, \url{https://youtu.be/YAaiDir7U2g}

Calibration videos: \url{https://youtu.be/W6-2dZ7TdmY}, \url{https://youtu.be/668u1TINfV0}

Single trial with tracking video: \url{https://youtu.be/lkfHRkv3YhI}

ElevenLabs: \url{https://elevenlabs.io/}

SVCL: \url{https://www.nirsoft.net/utils/sound_volume_command_line.html}

ARC-VR: \url{https://github.com/MPIB/arc-vr}

Meta Quest Pro: \url{https://www.meta.com/de/quest/quest-pro/}

ExtOSC: \url{https://github.com/Iam1337/extOSC}

Fraunhofer IDMT: \url{https://www.idmt.fraunhofer.de/}

Max-Planck Institute for Human Development: \url{https://www.mpib-berlin.mpg.de/en}

Unity: \url{https://unity.com/}

\section{Addendum: Voicelines Transcript}
\label{voicelines-transcript}

Below is a verbatim transcript of the voicelines used for the interactive tutorial (see section \ref{tutorial}). The respective sound-files were generated using the ElevenLabs API with the \verb|eleven_multilingual_v2| model using \verb|stlye=0|, \verb|stability=0.5| and \verb|similarity_boost=0.75| as parameters. A full video of the tutorial can be viewed here: \url{https://youtu.be/YAaiDir7U2g}.

\begin{lstlisting}

## Tutorial Voicelines

Sound/TutorialVoiceLines/welcome_message_en.mp3

Hello, and thank you for your participation in this experiment! Over the next few minutes, your task will consist of trying to locate sound-sources in three-dimensional space.


Sound/TutorialVoiceLines/barrier_disclaimer_en.mp3

You may move freely inside this rectangular area, highlighted in yellow. Just be sure to keep yourself and your hands within its bounds at all times. A yellow grid will appear if you get too close to the boundary.


Sound/TutorialVoiceLines/pause_disclaimer_en.mp3

If you feel unwell or tired, you can pause the experiment at any time and remove the VR headset. Refer to your experimenter for any further questions.


Sound/TutorialVoiceLines/before_we_begin_tutorial_en.mp3

Before we begin, we will walk you through an introductory test round.


Sound/TutorialVoiceLines/press_weiter_en.mp3

Press the "Continue" button to proceed.


Sound/TutorialVoiceLines/tutorial_you_will_hear_a_sound_en.mp3

Within the next few seconds, you will hear the ringing of a telephone. Try and locate the origin of this sound. Keep in mind that this point will be at head-height and within the yellow boundary.


Sound/TutorialVoiceLines/tutorial_select_sound_origin_en.mp3

Did you hear it? Move your hand to the location you think the sound originated from. Then, with your palm facing away from you, pinch your index-finger and thumb together to place a guess.


Sound/TutorialVoiceLines/tutorial_select_sound_origin_controller_en.mp3

Did you hear it? Move your controller to the location you heard the sound originate from. Then, with your index finger, press the highlighted trigger button to place a guess.


Sound/TutorialVoiceLines/tutorial_good_work_solution_en.mp3

Good work! Your selected location will appear with a yellow marker, whilst the actual sound-source is displayed as a red sphere. The closer these two are together, the better your score.


Sound/TutorialVoiceLines/tutorial_another_round_without_instructions_en.mp3

Next up, we will take you through a back-to-back set of three test rounds. This time, no instructions will be provided and, in addition to a ringing telephone, you may also hear the twirping of birds, or the sound of a piano. Just like before, try to locate the sound's origin and pinch your fingers at that location.


Sound/TutorialVoiceLines/begin_experiment_proper_en.mp3

Good work! With that, you are now ready to commence the experiment-proper. There will be a total of six blocks, each consisting of six back-to-back rounds just like the ones you completed just now, as well as two additional blocks of nine rounds, where the sound source will move. Ready?


Sound/TutorialVoiceLines/change_of_scene_en.mp3

The next six rounds will take place in a different virtual environment. The change will occur shortly.


Sound/TutorialVoiceLines/put_on_headphones_en.mp3

For the next set of rounds, please put on the stereo headphones attached to your VR headset. Once on, press the continue button to proceed.


Sound/TutorialVoiceLines/take_off_headphones_en.mp3

For the next set of rounds, please take off the stereo headphones attached to your VR headset. Once off, press the continue button to proceed.


Sound/TutorialVoiceLines/moving_sounds_en.mp3

For the next nine rounds, the sound-source will be moving along a line when it starts playing. Try and place your guess at the *endpoint*, the location at which it comes to a rest.


Sound/TutorialVoiceLines/experiment_ended_en.mp3

Good work! That concludes the experiment, and you may now take off the VR headset. Thanks a lot for participating, and have a nice day!
\end{lstlisting}

\end{document}